\documentclass[%
 reprint,
unsortedaddress,
 amsmath,amssymb,
prb,
]{revtex4-1}

\usepackage{graphicx}
\usepackage{dcolumn}
\usepackage{bm}

\usepackage{comment}
\usepackage{multirow}
\usepackage[english]{babel}
\usepackage[utf8]{inputenc}
\usepackage{ams math}
\usepackage{graphicx}
\usepackage[colorinlistoftodos]{todonotes}

\usepackage{dcolumn}
\usepackage{bm}
\usepackage{url}
\usepackage[colorlinks=true,linkcolor=blue,citecolor=blue,urlcolor=blue]{hyperref}

\usepackage[normalem]{ulem}

\begin{document}

\preprint{APS/123-QED}

\title{Screened-exchange density functional theory description of the electronic structure and phase stability of the chalcopyrite materials AgInSe$_2$ and AuInSe$_2$ }

\author{Namhoon Kim$^1$}
\author{Pamela Pe\~{n}a Martin$^2$}
\author{Angus A. Rockett$^{2,3}$}
\author{Elif Ertekin$^{1,3}$}
\affiliation{$^1$Department of Mechanical Science \& Engineering, 1206 W Green Street, University of Illinois at Urbana-Champaign, Urbana IL 61801}
\affiliation{$^2$Department of Materials Science \& Engineering, University of Illinois at Urbana-Champaign, Urbana IL 61801}
\affiliation{$^3$International Institute for Carbon Neutral Energy Research (WPI-I$^2$CNER), Kyushu University, 744 Moto-oka, Nishi-ku, Fukuoka 819-0395, Japan}

\email[e-mail: ]{ertekin@illinois.edu}

\date{\today}

\begin{abstract}
We present a systematic assessment of the structural properties, the electronic density of states, the charge densities, and the phase stabilities of AgInSe$_2$ and AuInSe$_2$ using screened exchange hybrid density functional theory, and compare their properties to those of CuInSe$_2$. 
For AgInSe$_2$, hybrid density functional theory properly captures several experimentally measured  properties, including the increase in the band gap and the change in the direction of the lattice distortion parameter $u$ in comparison to CuInSe$_2$. 
While the electronic properties of AuInSe$_2$ have not yet been experimentally characterized, we predict it to be a small gap ($\approx 0.15$ eV) semiconductor. 
We also present the phase stability of AgInSe$_2$ and AuInSe$_2$ according to screened-exchange density functional theory, and compare the results to predictions from conventional density functional theory, results tabulated from several online materials data repositories, and experiment (when available). 
In comparison to conventional density functional theory, the hybrid functional predicts phase stabilities of AgInSe$_2$ in better agreement with experiment: discrepancies in the calculated formation enthalpies are reduced by approximately a factor of three, from $\approx$ 0.20 eV/atom to $\approx$ 0.07 eV/atom, similar to the improvement observed for CuInSe$_2$. 
We further predict that AuInSe$_2$ is not a stable phase, and can only be present under non-equilibrium conditions.  
\end{abstract}

\pacs{Valid PACS appear here}
\maketitle


\section{\label{sec:level0} Introduction}

Chalcopyrite materials are important for a wide range of optoelectronic devices, for instance CuInSe$_2$ is well-known for its use as an absorber layer for applications in solar energy conversion. 
Thin-film photovoltaics based on the compound Cu(In,Ga)Se$_2$ (CIGS) have demonstrated the highest efficiencies to date among thin film polycrystalline materials~\cite{PIP:PIP2728}. 
In addition to CuInSe$_2$, other chalcopyrite materials such as AgInSe$_2$ have also recently become of interest~\cite{:/content/avs/journal/jvsta/30/4/10.1116/1.4728160}. 
Silver, which has the same number of valence electrons as copper, can be alloyed into CIGS in order to increase the band gap without causing structural disorder~\cite{Erslev20117296, Albornoz20141}. 
Although less studied to date, the chalcopyrite  AuInSe$_2$ has been suggested as a candidate material to observe a three-dimensional topologically insulating state \cite{PhysRevLett.106.016402, Feng2012}.
In addition, AuInSe$_2$ is reported to form during the vapor-liquid-sold growth of indium selenide nanowires from gold catalyst particles as well as possibly at the interface between gold/indium selenide metal-semiconductor junctions \cite{:/content/aip/journal/jap/54/10/10.1063/1.331808, :/content/aip/journal/jap/62/4/10.1063/1.339627, :/content/aip/journal/apl/89/23/10.1063/1.2388890}. 


However, while the properties of the chalcopyrite CuInSe$_2$ have received a lot of attention in the literature \cite{:/content/aip/journal/jap/108/2/10.1063/1.3456161, PhysRevB.87.245203, PhysRevLett.104.056401}, related  materials such as AgInSe$_2$ \cite{Sharma201497,C3EE41437J} and AuInSe$_2$ \cite{PhysRevLett.106.016402} have  received comparatively less.
These and other emerging materials may also be of interest for a variety of related applications, but their properties are less well understood. 
It is unknown how effective alloying AgInSe$_2$ into CuInSe$_2$ will be as a means to modify the band gap for photovoltaic applications and how the presence of Ag may affect transport or defect properties.  
Also, it is of interest to know what the properties of AuInSe$_2$ would be, if it were possible to synthesize the compound in the laboratory.  
Since there are no reports of the targeted synthesis of this compound, guidance on how it may be possible to synthesize the material may be useful. 
Therefore, in this work we present a comprehensive analysis of the electronic structure and phase stability of the selenide chalcopyrite materials AgInSe$_2$ and AuInSe$_2$ using screened exchange density functional theory. 

It is known from the literature that obtaining an accurate first-principles description of the electronic structure of CuInSe$_2$ is  challenging, due to the mixed ionic/covalent nature of the bonding and the need to properly describe the  $p$--$d$ hybridization in the solid \cite{PhysRevB.27.5176, PhysRevB.29.1882}. 
For instance, within the local density approximation or the generalized gradient approximation in density functional theory (DFT), the electronic band gap of CuInSe$_2$ (1.04 eV in experiment) closes.  
Hybrid density functionals have been shown to improve the description of the electronic structure and phase stability of CuInSe$_2$ \cite{:/content/aip/journal/jap/108/2/10.1063/1.3456161, PhysRevB.87.245203}, obtaining a better description of the electronic structure at a reasonable computational cost. 

Based on these considerations, we use hybrid screened-exchange DFT \cite{:/content/aip/journal/jcp/118/18/10.1063/1.1564060, :/content/aip/journal/jcp/121/3/10.1063/1.1760074, :/content/aip/journal/jcp/125/22/10.1063/1.2404663} and verify that it performs well on AgInSe$_2$ in addition to CuInSe$_2$ for electronic and thermodynamic properties.
The hybrid functional HSE06 correctly captures the increase in the band gap of AgInSe$_2$ compared to CuInSe$_2$, which is consistent with the predicted description of $p$-$d$ hybridization and structural parameters. 
We further demonstrate that the phase stability of AgInSe$_2$, calculated with the hybrid functional, compares well to experiment and offers similar improvements over conventional DFT based on PBE\cite{PhysRevLett.77.3865} as observed for CuInSe$_2$. 
Having verified the performance of the hybrid for the known chalcopyrites, we use it to predict the electronic and thermodynamic properties of AuInSe$_2$, which are as of yet unreported.
We predict AuInSe$_2$ to be a small gap semiconductor (in contrast to the predictions from data mining approaches\cite{Dey2014185}), which can only be present as a non-equilibrium phase. 
This suggests that synthesis of AuInSe$_2$ likely needs to proceed through non-equilibrium or potentially high pressure routes.

In Section II, we describe our methodology in detail. 
In Section III, we present the electronic structure of AgInSe$_2$ and AuInSe$_2$ in comparison to CuInSe$_2$. 
Section IV provides a discussion of phase stabilities of the compounds. 

\section{Methodology}

The DFT\cite{PhysRev.136.B864, PhysRev.140.A1133} calculations presented here were carried out within the Vienna {\it Ab Initio} Software Package (VASP)~\cite{Kresse199615, PhysRevB.54.11169}, using both the PBE~\cite{PhysRevLett.77.3865} and the HSE06~\cite{{:/content/aip/journal/jcp/125/22/10.1063/1.2404663}} approximations to the exchange-correlation potential.  
We used projector augmented wave (PAW) pseudopotentials~\cite{PhysRevB.50.17953, PhysRevB.59.1758} to represent the core electrons; we chose Ar, Kr, and Xe cores respectively for Cu, Ag, and Au, which keep the outer $d$ electrons in the valence.  
For each compound, the Kohn-Sham orbitals are expanded in a plane wave basis set with sufficient energy cutoff and $k$-point sampling of the Brillouin zone so that all computed parameters are converged to the number of significant figures shown. 
Total energies are reported for each compound at the DFT-optimized lattice constants according to each functional; when applicable internal atomic coordinates are relaxed so that all forces on each ion are $< 0.01$~eV/{\AA} for PBE as well as for HSE06. 

Hybrid functionals are a class of approximations to the unknown exchange-correlation energy functional that incorporate a portion $\alpha$ of exchange from Hartree-Fock theory. 
The HSE06 functional \cite{{:/content/aip/journal/jcp/125/22/10.1063/1.2404663}} used here uses a screened Coulomb potential to calculate the exchange. 
Although there have been attempts to match the band gap to the experimental value by adjusting the range separation parameter $\omega$ and/or exchange mixing parameter $\alpha$, in this work we set the simulation parameters to the suggested values $\alpha = 25\%$, $\omega = 0.2$ {\AA}$^{-1}$ for benchmarking. 
In the following, in addition to AgInSe$_2$ and AuInSe$_2$, for completeness and to facilitate comparison we also include results for CuInSe$_2$ (which have been previously reported \cite{:/content/aip/journal/jap/108/2/10.1063/1.3456161,PhysRevB.87.245203,0953-8984-23-42-422202}); our results for CuInSe$_2$ are in good agreement with others. 

The visualization program VESTA is used to analyze and visualize the computed electron density distribution \cite{Momma:db5098}.


\section{Structural Parameters and Electronic Structure}

\subsection{Structural Parameters}

The chalcopyrite structure $ABC_2$ (space group I$\overline{4}$2d)  is closely related to the zinc blende lattice typical of binary semiconductors. 
However, two cation species ($A,B$) are present in an ordered arrangement on the usual cation sublattice. 
As a result, each anion $C$ has two $A$ and two $B$ cations as nearest neighbors, which leads to a symmetry breaking tetragonal distortion. 
The conventional unit cell of the chalcopyrites is thus double the conventional unit cell of the zinc blende lattice, with two of the latter stacked together and $c \ne 2a$. 
Additionally, the anion $C$ typically assumes an equilibrium position closer to one pair of cations than to the other (bond lengths are unequal).
Therefore, the $X$InSe$_2$ ($X$ = Cu, Ag, Au) chalcopyrites  can be characterized by three structural parameters: the lattice constants $a$ and $c$ and the anion displacement $u$. The $u$ parameter represents how the anion, Se, is displaced from its ideal tetrahedral site:
\begin{equation}
u = \frac{1}{4} + \frac{d^2_{X-Se}-d^2_{In-Se}}{a^2} 
\hspace{1em},
\end{equation}
where $d_{X-Se}$ and $d_{In-Se}$ are the bond length of $X$-Se and In-Se, respectively. The value $u =$ 0.25 for tetrahedral bonding, with $u < 0.25$  denoting displacement towards the $X$ atom. 
Since $u$ is related to the degree of orbital hybridization between different atoms, many calculated properties such as band gap are sensitive to it\cite{PhysRevB.27.5176, PhysRevB.29.1882, PhysRevLett.104.056401}.

\begin{table*}
\caption{ \label{lattice} Calculated structural parameters $a$, $c/a$ and $u$, and band gaps of $X$InSe$_2$ ($X$ = Cu, Ag, Au). 
Percent differences compared to experimental data (when available) are shown in parentheses. }
\begin{ruledtabular}
\begin{tabular}{ccccccccccccc}
&\multicolumn{3}{c}{$a$ ({\AA})} &\multicolumn{3}{c}{$c/a$ ({\AA})} &\multicolumn{3}{c}{$u$} &\multicolumn{3}{c}{Band gap (eV)} \\ \cline{2-4} \cline{5-7} \cline{8-10}  \cline{11-13}
& PBE & HSE06 &Exp. & PBE & HSE06 &Exp. & PBE & HSE06 &Exp. & PBE & HSE06 &Exp. \\ \hline 
\multirow{2}{*}{CuInSe$_2$} 
& 5.88 & 5.83  & 5.78\footnotemark[1] & 2.01 & 2.01 & 2.01\footnotemark[1]  & 0.22 & 0.23 & 0.23\footnotemark[1] & 0 & 0.83 & 1.04\footnotemark[3]  \\
& (1.8) & (0.8) & & (-0.1) & (-0.2) & & (-3.7) & (0.3) & &   & (-19.8)  & \\ \hline 
\multirow{2}{*}{AgInSe$_2$} 
& 6.20 & 6.16 & 6.10\footnotemark[2] & 1.95 & 1.93 & 1.92\footnotemark[2] & 0.25 & 0.26 & 0.26\footnotemark[2] & 0 & 0.97 & 1.24\footnotemark[3] \\ 
& (1.6) & (1.0) & -- & (1.6) & (0.8) & -- & (-2.4) & (0.3) & -- & & (-22.1) & \\  \hline 
AuInSe$_2$ 
& 6.14 & 6.07 & -- & 1.99 & 1.98 & -- & 0.24 & 0.25 & -- & 0 & 0.16 & --  \\
\end{tabular}
\end{ruledtabular}
\footnotetext[1]{Reference~\cite{KNIGHT1992161}}
\footnotetext[2]{Reference~\cite{1347-4065-19-S3-85}}
\footnotetext[3]{Reference~\cite{SHAY1975110}}
\end{table*}

The structural parameters for $X$InSe$_2$, ($X=$ Cu,Ag,Au) calculated by DFT using both the PBE and HSE06 and the corresponding experimental values are summarized in Table~\ref{lattice}. 
For $X=$ Cu and Ag, the lattice constants $a$ and $c$ estimated by PBE are larger than experimental data, showing deviations of around 2\% from the experimental value. 
HSE06 improves $a$ and $c$, bringing the values closer to experiment (within 1\%). 
For CuInSe$_2$, our calculated PBE and HSE06 parameters are in agreement with previously reported values\cite{{:/content/aip/journal/jap/108/2/10.1063/1.3456161}, {PhysRevB.87.245203}}.
For AgInSe$_2$, our PBE results are consistent with previously reported values \cite{Sharma201497}.

For AgInSe$_2$ the description of the distortion $u$ is also improved in HSE06, similar to the improvement observed for CuInSe$_2$.
According to Table~\ref{lattice}, the deviation of $u$ from experiment is large in PBE ($\approx$ 3\%) but HSE06 improves it to within 0.3\%. 
Notably HSE06 properly captures the trend $u < 0.25$ (Se atom shifts towards Cu) for CuInSe$_2$ \cite{KNIGHT1992161} but $u > 0.25$ (Se atom shifts away from Ag) in AgInSe$_2$ \cite{1347-4065-19-S3-85}. 
The difference in the direction of the distortion may be due to the smaller size of Cu atoms compared to Ag atoms and/or a stronger interaction matrix element betweeen the Cu-$s$/Se-$p$ core states in comparison to Ag-$s$/Se-$p$. 
Since $X$--Se spacing is related closely to the degree of hybridization between the transition metal $d$ orbitals and the $4p$ orbitals of the chalcogen, it is encouraging that the match to experiment improves. 

Table~\ref{lattice} also provides the calculated structural parameters for AuInSe$_2$ according to PBE and HSE06. 
While we cannot compare it to experimental data, it would be of interest to know whether as in the cases of $X$=Cu, Ag the use of the hybrid improves the description, or whether for $X$=Au the system is sufficiently delocalized that the PBE description is more appropriate. 
Based on the calculated parameters, we predict that the lattice constants of AuInSe$_2$ lie in between those of CuInSe$_2$ and AgInSe$_2$, and that $u \approx$ 0.25, suggesting less symmetry breaking and a more tetrahedral structure. 

\subsection{Electronic Structure and Density of States}

\begin{figure}
\includegraphics{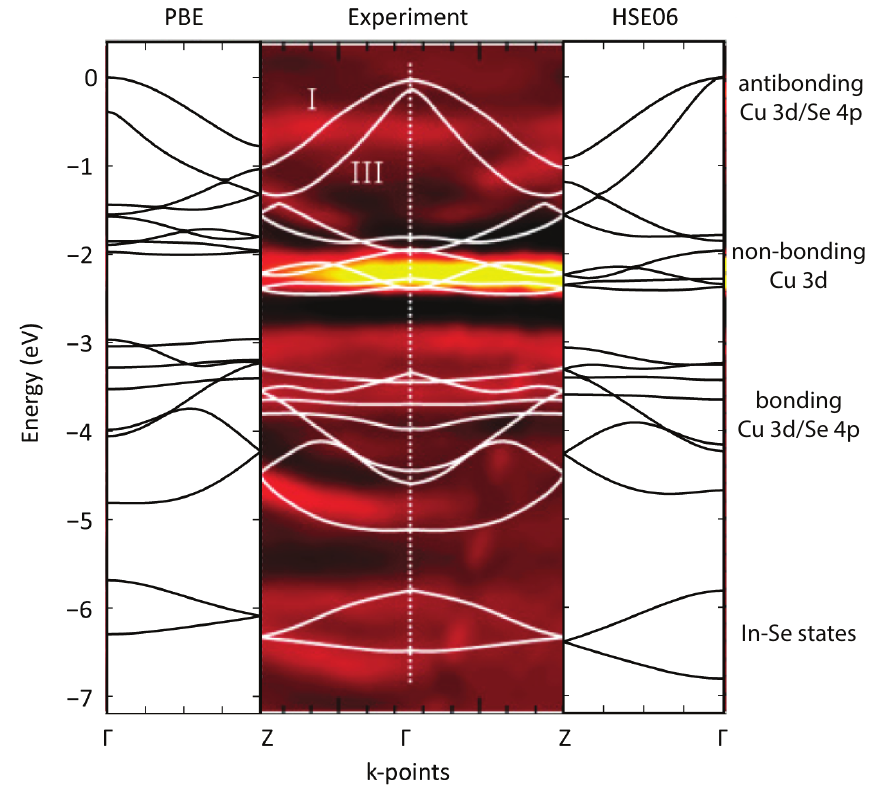} 
\caption{\label{band}
Electronic band structure of CuInSe$_2$ according to DFT-PBE (left) and DFT-HSE06 (right), in comparison to recent ARPES measurements (middle).  Adapted with permission from Ref. [\onlinecite{PhysRevB.84.115109}]. Copyrighted by the American Physical Society.}
\end{figure}

\begin{figure*}
\includegraphics{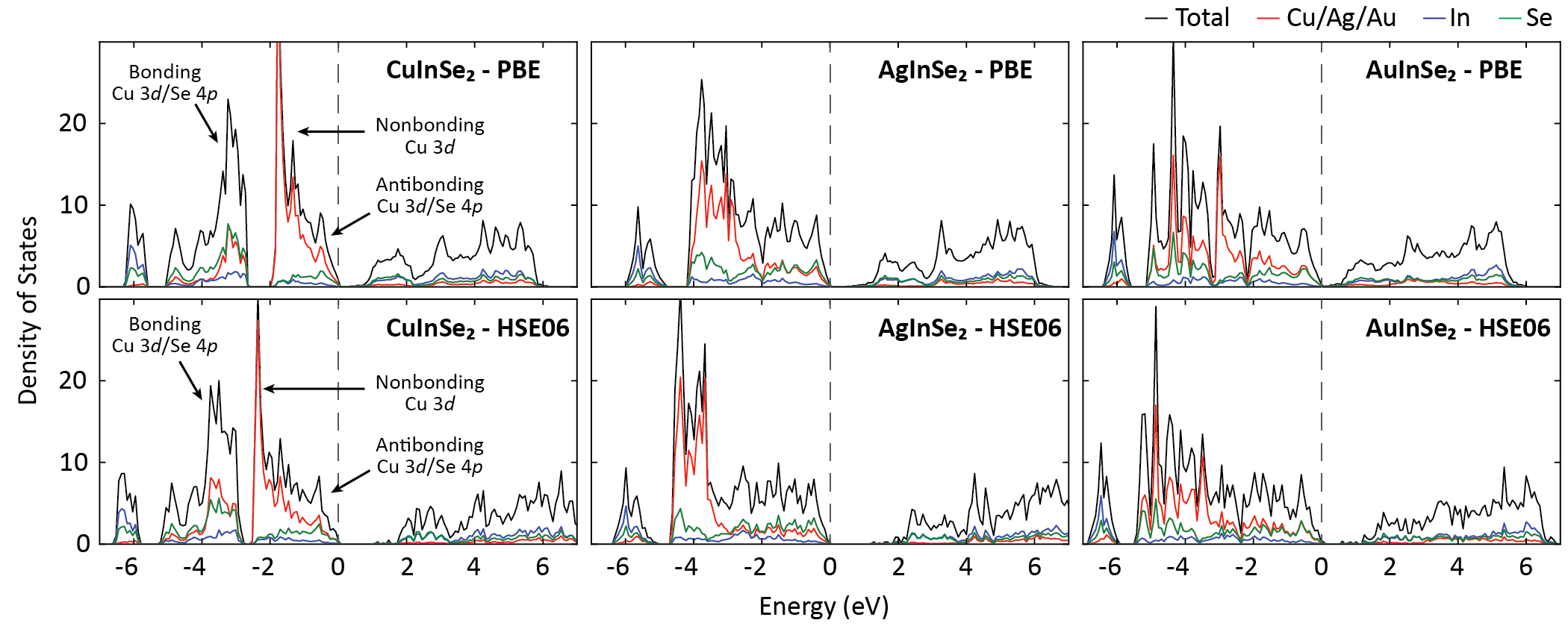}
\caption{\label{dos}
Density of states (DOS) for CuInSe$_2$ (left), AgInSe$_2$ (middle), AuInSe$_2$ (right) calculated by DFT-PBE (upper) and DFT-HSE06 (lower). 
The black lines are total DOS, and red, blue and green lines are projected DOS of Cu/Ag/Au, In, and Se respectively.
The Fermi level is shown by the dashed grey line.}
\end{figure*}

In comparison to the corresponding binary II-VI semiconductors, the chalcopyrites exhibit a suppressed band gap: for instance in CuInSe$_2$ the gap is $E_g = 1.04$~eV \cite{SHAY1975110}, much lower than its II-VI counterpart ZnSe for which $E_g = 2.68$ eV \cite{SHAY1975110}.
The reason for the suppressed gap can be understood from Fig.~\ref{band}, which shows the structure of CuInSe$_2$ valence bands according to PBE and HSE06, in comparison to recent angle-resolved photoemission spectroscopy (ARPES) measurements~\cite{PhysRevB.84.115109}. 
The differences between the PBE and HSE06 band structure of CuInSe$_2$ have been noted previously \cite{0953-8984-23-42-422202}; we summarize the key concepts here to observe similarities and differences with AgInSe$_2$ and AuInSe$_2$. 
The projected density of states (PDOS) of the CuInSe$_2$ valence bands onto the atomic states are also shown in Fig.~\ref{dos} (leftmost column), which show the nature of the electronic structure as (1) uppermost VBs: antibonding Cu $3d$/Se $4p$ states, (2) ~2 eV below VBM: a set of non-bonding Cu 3$d$ states, and (3) ~4 eV below VBM: the bonding Cu $3d$/Se $4p$ levels, and (4) ~6 eV below VBM: bands derived from interactions between In and Se atoms. 
According to Fig.~\ref{band}, for CuInSe$_2$, HSE06 improves the orbital spacing and bandwidths of the valence bands in comparison to PBE (the relative positions are better matched to experiment).
The calculated band structures for AgInSe$_2$ and AuInSe$_2$ (not shown) are qualitatively similar to that of CuInSe$_2$ in Fig.~\ref{band} in terms of the orbital ordering, but the orbital spacing and bandwidths vary.

For CuInSe$_2$, interactions between the non-bonding Cu 3$d$ states and the antibonding Cu 3$d$/Se 4$p$ levels towards the top of the valence bands in Fig.~\ref{band} leads to a repulsive $p$--$d$ interaction that pushes the antibonding states upwards in energy, which is largely responsible for the reduced band gap \cite{PhysRevB.27.5176, PhysRevB.29.1882}.
Capturing this interaction properly is necessary to accurately position the bands and obtain a good estimate of the band gap.
For instance, for CuInSe$_2$ (Fig.~\ref{band}) according to PBE the $3d$ non-bonding levels are too high in energy, resulting in an ``overhybridization" with the upper valence bands. 
The overhybridization pushes the upper VBs further upwards and is largely responsible for the closure (in fact, the inversion\cite{inversion}) of the gap in PBE, see Fig. \ref{dos}. 
The use of HSE06 drops the position of Cu $3d$ non-bonding orbitals and opens the gap to $E_g^{HSE06} = 0.83$ eV, in better agreement with the experimental value of 1.04 eV \cite{SHAY1975110}. 

To see what changes in the electronic structure of $X$InSe$_2$ for $X$=Ag, Au in comparison to $X$=Cu, the total and projected density of states (DOS) of the $X$InSe$_2$ compounds are shown in Fig.~\ref{dos}.
The dotted line indicates the top of the valence band in all cases. 
The black lines are the total DOS, and red, blue, and green indicate the transition metal $X$, In, and Se states respectively. 
For AgInSe$_2$ the first observation is that the HSE06 band gap is larger than the PBE gap (as expected).
In PBE the gap is closed (in fact, inverted), while in HSE06 it opens to 0.97 eV. 
The opening of the gap in AgInSe$_2$ in HSE06 occurs for similar reasons as discussed above for CuInSe$_2$: the DOS in Fig.~\ref{dos} (middle column)  shows that the nonbonding 4$d$ levels are shifted downwards in HSE06, reducing the $p$ -- $d$ repulsion, and opening the gap.    
In addition, according to HSE06 the gaps are $E_g = 0.83$ eV for CuInSe$_2$ and $E_g = 0.97$ eV for AgInSe$_2$, in reasonable agreement with the experimental values (1.04 eV and 1.24 eV respectively).
Therefore, HSE06 is also able to capture the increase in the gap when Cu is replaced with Ag (discussed further later). 

The encouraging performance of the hybrid functional for AgInSe$_2$ as well as CuInSe$_2$ suggests that it may be well-suited for AuInSe$_2$ as well. 
Interestingly we find that the trend described above persists somewhat for AuInSe$_2$: the gap is closed (inverted) in PBE, but $E_g \approx 0.16$ eV in HSE06. 
Thus, although the gap opens in HSE06, it does so to a lesser extent than the other chalcopyrites. 
To understand why, from Fig.~\ref{dos} as $X$ changes from Cu $\rightarrow$ Ag $\rightarrow$ Au, the relevant $d$ orbitals become more delocalized in nature and exhibit a larger bandwidth in the DOS. 
Accordingly the degree of mixing between the non-bonding $X$-$d$ and antibonding $X$-$d$/Se 4$p$ states within the upper valence bands increases. 
The more delocalized nature of the 5$d$ bands is already better described in PBE, compared to the more localized 3$d$ and 4$d$ bands.   
Correspondingly, when HSE06 is used in AuInSe$_2$, there is no need for as large a ``correction" to the position of the  nonbonding Au $5d$ states and the HSE06 and PBE DOS (and thus, the gap) for AuInSe$_2$ are more similar.  

 
\begin{figure*}
\includegraphics{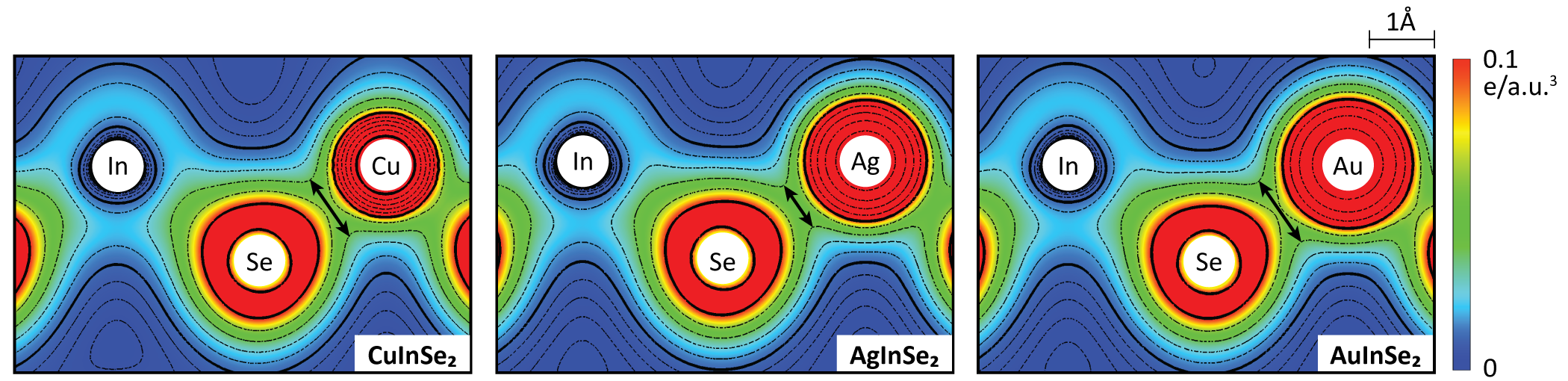}
\caption{\label{charge}
Electronic charge density distribution of CuInSe$_2$ (left), AgInSe$_2$ (middle), AuInSe$_2$ (right), across a plane that contains Cu/Ag/Au, In and Se atoms according to DFT-HSE06. The contour lines of isosurfaces are spaced logarithmically and the arrows are marked corresponding to the same isosurface value for all three materials.}
\end{figure*}


Overall the HSE06 gap of AgInSe$_2$ (0.97 eV) is larger than that of CuInSe$_2$ (0.83 eV), and then quickly drops to 0.16 eV for AuInSe$_2$.  
The increase in the gap as Cu $\rightarrow$ Ag is consistent with experiment, but the non-monotonic trend as Cu $\rightarrow$ Ag $\rightarrow$ Au is different from typical expectations of a reduced gap as isovalent elements in a compound are successively exchanged by descending along a column of the periodic table~\cite{doi:10.1021/ed069p151}.  
These trends can be better understood from the calculated distortion parameter $u$ in Table \ref{lattice} and the HSE06 charge densities shown in Fig.~\ref{charge}, plotted on a plane that contains $X$, In, and Se atoms. 
The scale bar and the color map isosurface levels are the same for all three plots, for direct comparison of the charge distribution in the three materials.

According to Fig.~\ref{charge}, both ionic and covalent bonds are present along the In-Se and $X$-Se bonds, respectively, in the $X$InSe$_2$ system. 
Electrons are drawn towards Se and away from In in the In-Se bond, reflecting a more ionic nature, while the more evenly distributed charge in the $X$-Se bond indicates a more covalent nature. 
For the $X$-Se bond, arrows indicating the width corresponding to a particular isosurface value for all three materials are marked in Fig.~\ref{charge}. 
The arrow is widest for AuInSe$_2$, narrowest for AgInSe$_2$, and intermediate for CuInSe$_2$. 
This ordering is consistent with the relative band gaps in the compounds, as well as the differences in the $u$ parameter and varying degrees of $p$ -- $d$ hybridization for all materials in Fig. \ref{dos}. 
As $u < 0.25$ in CuInSe$_2$, the Cu and Se atoms are closer together, and orbital overlap and $p$-$d$ repulsion is greater in this material, resulting in a smaller band gap.  
By contrast, for AgInSe$_2$ where $u > 0.25$, the Se atom is further away from Ag and closer to In. 
Less direct orbital overlap and $p$-$d$ repulsion within the upper valence bands results in a larger gap for this material.
For AuInSe$_2$, $u =$ 0.25 suggests a more typical tetrahedral bonding. 
The high degree of charge sharing in the AuInSe$_2$ results from the larger atomic size of Au and the 6$d$ orbitals, and the more delocalized nature of these orbitals results in smaller $p$ -- $d$ repulsion and generally more delocalized states and metallic like character in the solid.


\section{Phase Stabilities and Accuracy Assessment}

Having characterized the electronic structure of the materials, we now turn to the formation enthalpies and phase stabilities.  
Not only is this analysis directly of relevance for the chalcopyrites themselves, but it also suggests the degree of accuracy attainable in the calculation of phase stabilities in general and thus has implications for the long-standing goal of computational materials prediction and discovery \cite{Asta2014}. 
Therefore, in the following the phase stability according to PBE and HSE06 are compared to each other and experiment (when available). 
Results for CuInSe$_2$ phase stability have been previously reported\cite{:/content/aip/journal/jap/108/2/10.1063/1.3456161, PhysRevB.87.245203}, we show them here again for comparison purposes. 
For completeness, we also compare our computed phase stabilities to those that are available in several online materials repositories \cite{Jain2013,PhysRevB.85.115104,Saal2013} to assess patterns and trends in the results.  



\subsection{Formation Enthalpies}

In a thermodynamic framework, phase stability diagrams determine the regions of chemical potential phase space for which a particular compound is stable with respect to the formation of all possible elemental solids and other binary or ternary compounds. 
Different growth or environmental conditions are accommodated by the set of chemical potentials $\mu_i$ for each atomic species $i$ that is present in the compound, by assuming that each is in equilibrium with a physical reservoir such as a gas or a bulk phase. 
At thermodynamic equilibrium the chemical potentials satisfy
\begin{equation}
\mu_{comp} = \sum_i \mu_i \hspace{1em},
\end{equation}
where the summation is carried out over all atomic species present in the compound, {\it e.g.} $\mu_{CuInSe_2} = \mu_{Cu} + \mu_{In} + 2 \mu_{Se}$. 
By referencing the chemical potential of each atomic species in the compound to that of the bulk elemental solid or gas phase so that $\mu_i = \mu^o_i + \Delta \mu_i$, then  
\begin{equation} 
\mu_{comp} = \sum_i \mu_i^o + \sum_i \Delta \mu_i \hspace{1em}.
\end{equation}
The formation enthalpy of the compound, 
\begin{eqnarray} 
\Delta H & = & \sum_i \Delta \mu_i \label{eq:enthalpy}\hspace{1em}, 
\end{eqnarray}
is the change of enthalpy when the compound is formed from its constitutent elemental phases. 
Negative $\Delta H$ denotes stable compounds. 
Note that we neglect TS and PV contributions to the enthalpy here. 

\begin{table}[b] 
\caption{\label{enthalpy}%
Calculated enthalpy of formation of $X$InSe$_2$ (X = Cu, Ag, Au) per formula unit. Percent differences compared to experimental data are shown in parentheses. }
\begin{ruledtabular}
\begin{tabular}{ccccccc}
& \textrm{PBE} & \textrm{HSE06} & Mat Proj\cite{Jain2013} & NREL\cite{PhysRevB.85.115104} & OQMD\cite{Saal2013} &  \textrm{Exp} \\ \colrule
\multirow{2}{*}{CuInSe$_2$} 
& -1.78 & -2.40 & -1.77 & -2.32 & -1.80 & -2.77\footnotemark[1] \\
& (-35.6) & (-13.4) & (-36.1) & (-16.3) & (-35.0) & \\
\multirow{2}{*}{AgInSe$_2$} 
 & -1.71 & -2.22 & -1.72 & -2.08 & -1.73 &  -2.51\footnotemark[1] \\
& (-31.8) & (-11.4) & (-31.5) & (-17.1) & (-31.1) & \\
AuInSe$_2$ 
& -1.23 & -1.59  & -- & -- & -- & -- \\
\end{tabular}
\end{ruledtabular}
\footnotetext[1]{Reference~\cite{formE_CIS_AIS}}
\end{table}

The calculated formation enthalpies of the chalcopyrite materials according to PBE and HSE06 are shown in Table~\ref{enthalpy}, in comparison to the experimental values.  
For AgInSe$_2$, the enthalpy of formation from PBE is underestimated by more than 30\% of the experimental values (similar to the underestimate observed for CuInSe$_2$). 
The underestimate occurs because the elemental compounds Cu or Ag, In, and Se are better described in PBE than the chalcopyrites themselves.  
Since the chalcopyrites and the elemental solids exhibit different degrees of metallic, covalent, or ionic character, PBE suffers from an incomplete cancellation of errors.  
The incomplete cancellation of errors arises because PBE is not able to describe materials with different degrees of electron localization on equal footing~\cite{PhysRevB.85.115104, PhysRevB.84.045115, PhysRevB.85.155208}. 

According to Table~\ref{enthalpy}, HSE06 also underestimates the formation enthalpy of AgInSe$_2$, but now the error is reduced to $\approx$ 11\%. 
Therefore, it appears that HSE06 improves the description overall (and the level of improvement is quite similar to that observed for CuInSe$_2$).  
For comparison, we also show the formation energies according to NREL Materials Database\cite{PhysRevB.85.115104}, the Open Quantum Materials Database (OQMD)\cite{Saal2013}, and Materials Project\cite{Jain2013}.
The results from these online repositories mostly match well our PBE calculations, as expected since similar calculation protocols are used.  
For example, the Materials Project uses DFT to evaluate the total energy of compounds and mixes GGA and GGA+U for the exchange-correlation functional\cite{PhysRevB.84.045115} for accurate prediction of formation enthalpies. 
The NREL Materials Database formation enthalpy is different from the other two, as a result of the {\it Fitted Elemental Reference Energies} (FERE\cite{PhysRevB.85.115104}) approach. 
Interestingly in this case this brings the formation enthalpies more into accord with our hybrid DFT results and closer to the experimental values. 

For AuInSe$_2$, we predict formation enthalpies of -1.23 eV/fu in PBE and -1.59 eV/fu in HSE06; results for this material are not available in the repositories or experiment. 
This suggests that the compound is favorable to form from isolated elemental solids Au, In, and Se.  
However, the formation enthalpy is lower than the predicted value for the other chalcopyrite materials. 
For this compound, we note that the PBE and HSE06 description are more similar to each other on an absolute scale, in comparison to the other two chalcopyrites. 

\subsection{Formation Enthalpies of Competing Binary Phases} 

\begin{figure*}
\includegraphics{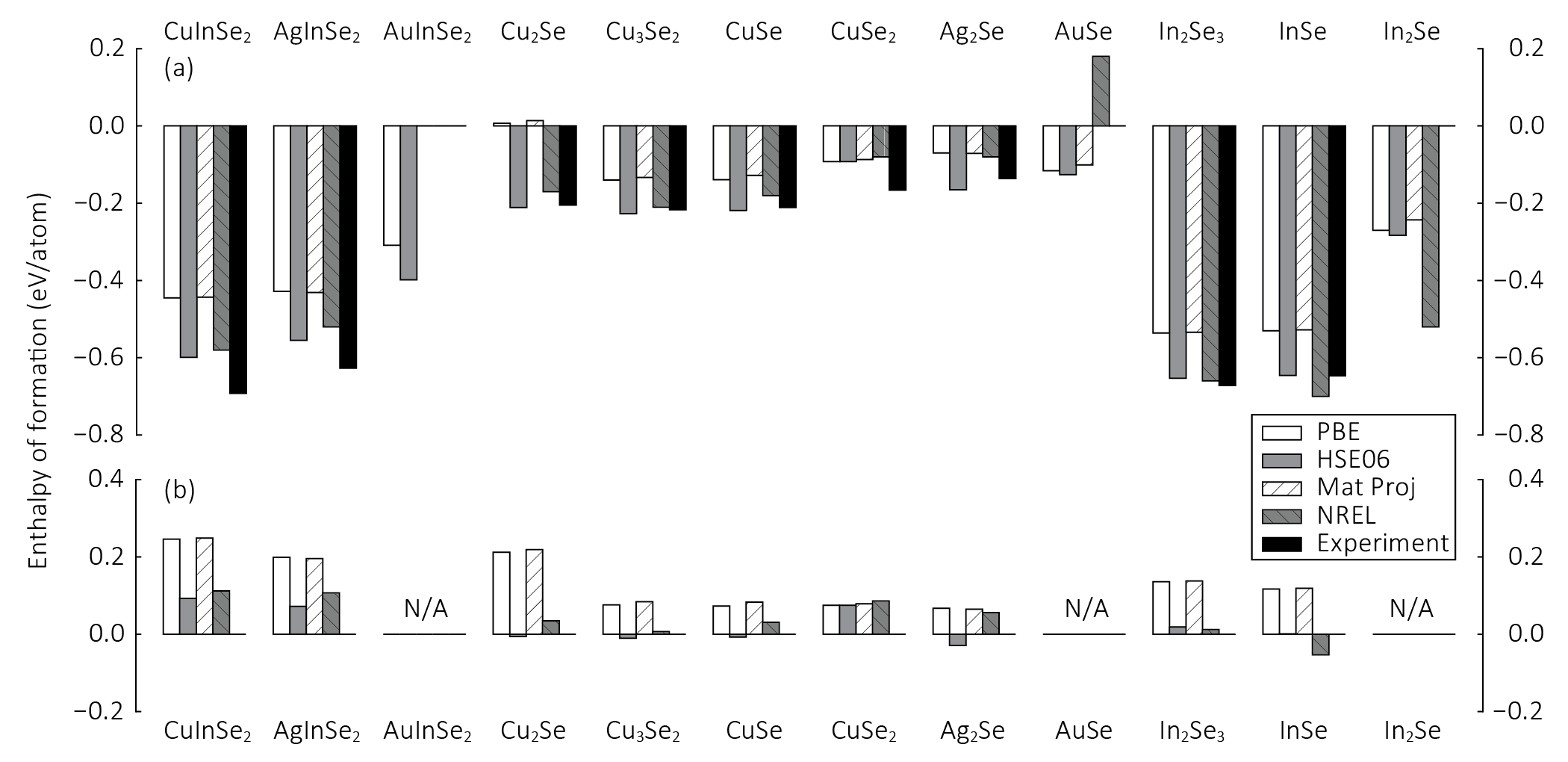}
\caption{\label{formation}
Comparisons of formation enthalpies of the chalcopyrites and the binaries calculated by DFT-PBE and DFT-HSE06 with experimental data for  $X$InSe$_2$~\cite{formE_CIS_AIS}, Cu-Se and In-Se binaries~\cite{NREL_report}, and Ag-Se binary~\cite{NBS_tables}. (a) Absolute values of formation enthalpies in eV per atom.  (b) Absolute deviations from the experimental values in eV/atom.}
\end{figure*}

A detailed comparison of formation enthalpies of the chalcopyrites and several competing candidate binary phases is shown in in Fig.~\ref{formation}. 
Fig.~\ref{formation}a shows absolute values of formation enthalpies, and Fig.~\ref{formation}b shows the absolute deviations from the experimental values, both on a per atom basis. 
The compounds shown in Fig.~\ref{formation} were selected by using the known experimental formation enthalpies as a guide to identify the most competitive alternate phases. 
For CuInSe$_2$, Cu-Se binaries such as Cu$_2$Se, Cu$_3$Se$_2$, CuSe, CuSe$_2$ were selected, and for AgInSe$_2$, Ag$_2$Se was included. 
The In-Se binaries, In$_2$Se$_3$, InSe, In$_2$Se were incorporated in the phase space for both cases. 
Some of these phases are metallic (In$_2$Se and all of the copper selenides except for Cu$_2$Se) and others semiconducting (the chalcopyrites, Cu$_2$Se, Ag$_2$Se, AuSe, In$_2$Se$_3$, and InSe), see the DOS according to HSE06 presented in the Appendix.
In some cases in Fig.~\ref{formation} (AuInSe$_2$, AuSe, and In$_2$Se) the experimental enthalpies of formation are not included because we are unable to find reported values. 

Several trends are evident from Fig.~\ref{formation}.  
First, as expected, it is clear that PBE underbinds the compounds with respect to the solid elemental phases. 
On a per atom basis, the worst cases are the chalcopyrites CuInSe$_2$ and AgInSe$_2$ themselves and the semiconductor Cu$_2$Se for which the errors are all larger than 0.20 eV/atom. 
In fact, Cu$_2$Se is predicted to be slightly unstable in PBE. 
Based on their density of states (see Appendix), these worst cases correspond to semiconductors for which the valence bands have a more substantial $d$ orbital character. 
For these materials, HSE06 brings the formation enthalpy closer to the experimental value although it still underbinds (after all, ultimately, HSE06 is still not self-interaction free). 
For the chalcopyrites, the degree of underbinding is reduced to  0.07-0.09 eV/atom while for Cu$_2$Se the error is reduced to within 0.01 eV/atom in comparison to experiment. 

For the remaining semiconductors Ag$_2$Se, In$_2$Se$_3$, and InSe where the valence bands have less $d$ orbital character, both PBE and HSE06 both do better.  
In these cases, the PBE results are closer to experiment to begin with, exhibiting deviations from experiment of 0.07 eV/atom, 0.14 eV/atom, and 0.12 eV/atom respectively.  
Here again, HSE06 results improve the description further, bringing the deviation to within 0.03 eV/atom, 0.02 eV/atom, and 0.01 eV/atom respectively.  

Considering the metallic compounds Cu$_3$Se$_2$, CuSe, and CuSe$_2$, the PBE description is better as expected: the formation enthalpy is always within $\approx$ 0.07 to 0.08 eV/atom in all cases.
In the case of Cu$_3$Se$_2$ and CuSe, HSE06 improves the description to within 0.01 eV/atom, but it does not change the description of CuSe$_2$. 

For comparison purposes, in Fig. \ref{formation} we also show the formation enthalpies according to Materials Project and the NREL database.  
As expected the Materials Project results agree well with our PBE results. 
It is interesting to observe that, with the exception of AuSe for which there is a large deviation, for most of these compounds the NREL Materials Database results, obtained using FERE\cite{PhysRevB.85.115104}, match reasonably well the results of the hybrid calculations presented here. 


\subsection{Phase Stability Diagrams} 

\begin{figure*}
\includegraphics{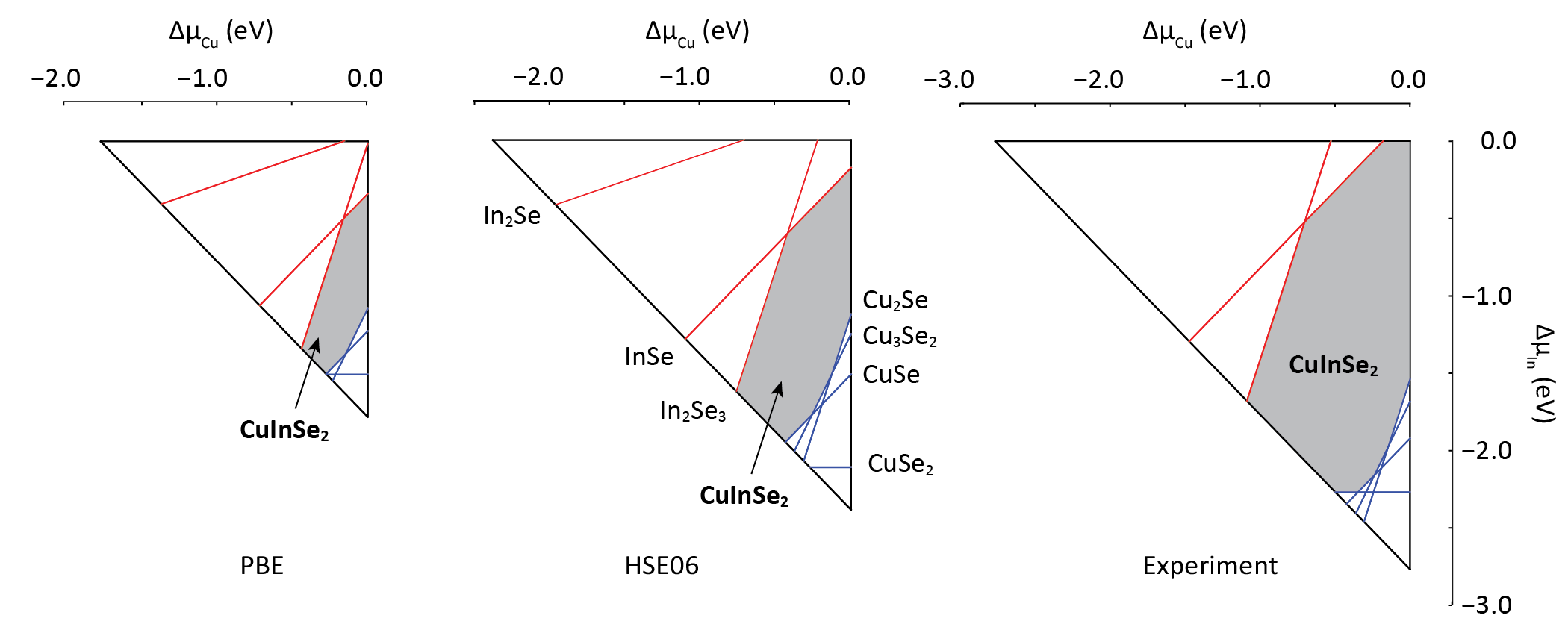}
\caption{\label{cuinse2}
Phase stability diagrams of CuInSe$_2$ according to DFT-PBE (left), DFT-HSE06 (middle), and experimental data (right). The area below the blue lines or above the red lines indicate where CuInSe$_2$ is unstable with respect to formation of Cu-Se or In-Se binaries, respectively. The shaded area shows the chemical potential range for which CuInSe$_2$ is stable.}
\end{figure*}

\begin{figure*}
\includegraphics{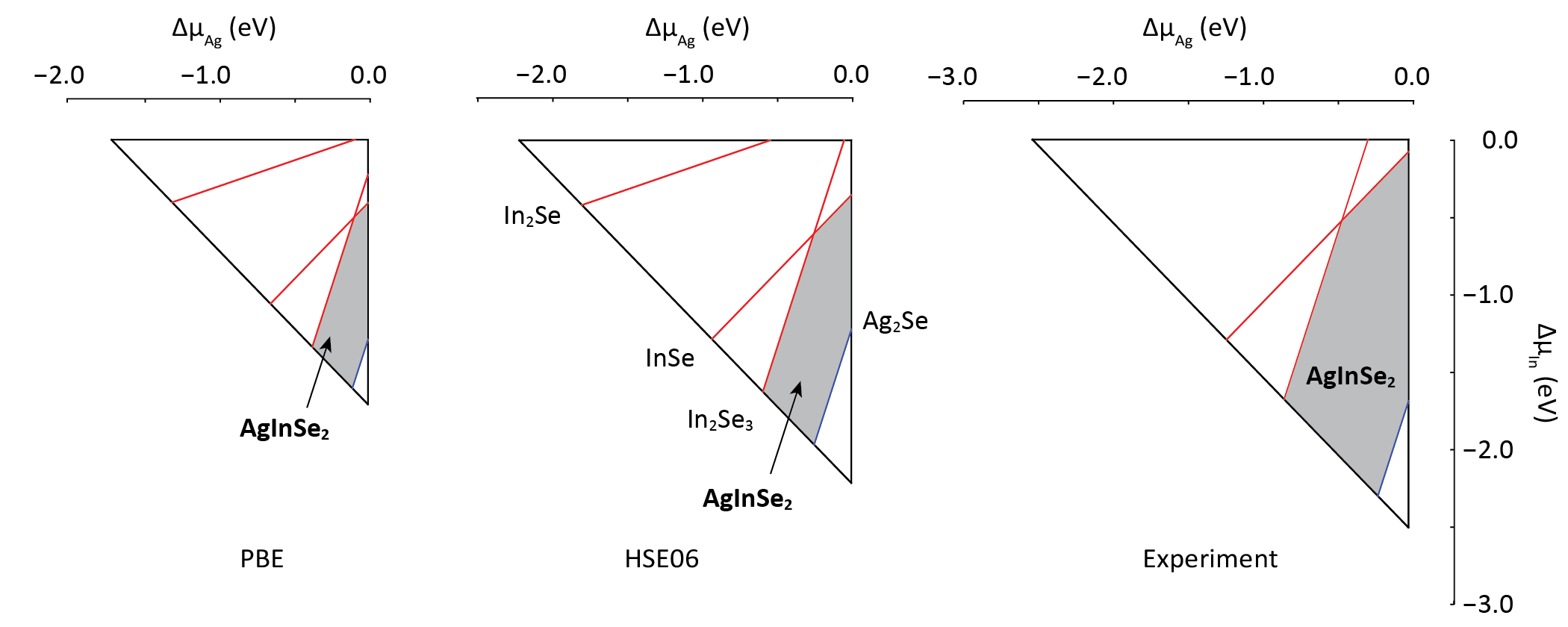}
\caption{\label{aginse2}
Phase stability diagrams of AgInSe$_2$ according to DFT-PBE (left), DFT-HSE06 (middle), and experimental data (right). The area below the blue line or above the red lines indicate where AgInSe$_2$ is unstable with respect to formation of Ag-Se or In-Se binaries. The shaded area shows the chemical potential range for which AgInSe$_2$ is stable.}
\end{figure*}

\begin{figure}
\includegraphics{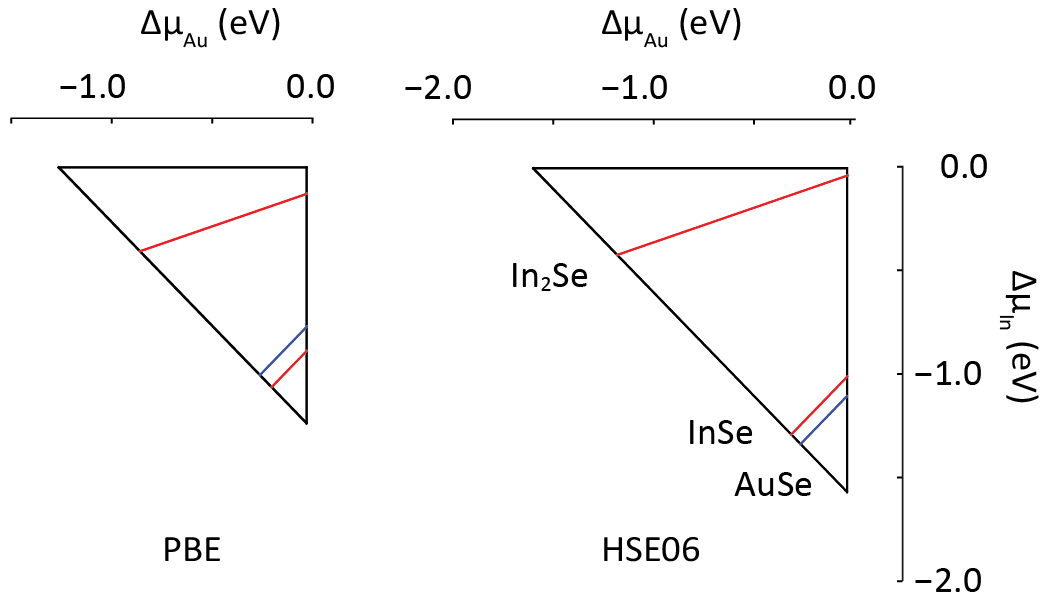}
\caption{\label{auinse2}
Phase stability diagrams of AuInSe$_2$ from DFT-PBE (left), DFT-HSE06 (right). The area below the blue lines or above the red lines indicate where Au-Se or In-Se binaries are stable than AuInSe$_2$. AuInSe$_2$ is not stable in either DFT functional for any value of chemical potentials.}
\end{figure}

Finally, phase stability diagrams for CuInSe$_2$, AgInSe$_2$, and AuInSe$_2$ obtained from the chalcopyrite and competing phase formation enthalpies are shown in Figs.~\ref{cuinse2},  \ref{aginse2}, and Fig.~\ref{auinse2}. 
For CuInSe$_2$ and AgInSe$_2$ we compare PBE and HSE06 calculation results to experimental values directly; for AuInSe$_2$ the experimental values are not known so only calculation results are presented. 
Each point inside the triangles corresponds to a set of chemical potentials satisfying thermodynamic equilibrium, {\it e.g.} $\Delta \mu_{Cu} + \Delta \mu_{In} + 2\Delta \mu_{Se} = \Delta H_{CuInSe_2} $, and the constraint that $\Delta \mu_i \leq 0$ so that the compound is not unstable with respect to decomposition to elemental solids.  
This gives rise to the outer triangles in Figs.~\ref{cuinse2},  \ref{aginse2}, and \ref{auinse2}, the size of which is governed by the value of $\Delta H$ for the chalcopyrites. 
The shaded region corresponds to the part of phase space for which the chalcopyrite is also stable with respect to decomposition to competing binaries $\sum_i \Delta \mu_i \leq \Delta H_{binary} $.

The trends observed for AgInSe$_2$ in Fig. \ref{aginse2} are quite similar to those of CuInSe$_2$ in Fig. \ref{cuinse2}. 
Since from Table~\ref{enthalpy}, the formation enthalpy in PBE is small compared to experiment, the size of the PBE triangle is small but the HSE06 triangle is larger and closer to experiment.  
Boundaries marking where the compound is unstable with respect to a particular binary are represented as lines in Figs.~\ref{cuinse2} and \ref{aginse2}, with blue lines representing $X$-Se binaries and red lines representing In-Se binaries. 
It is interesting that even in PBE, despite the smaller size of the triangles and the large degree of underbinding of AgInSe$_2$, the compound is still predicted to be stable in some portion of phase space (the same holds for  CuInSe$_2$).  
We note that this may not always be the case: it is conceivable that if the degree of underbinding were slightly larger, the compound may be incorrectly predicted to be unstable (a ``false negative"). 
This possibility highlights some of the challenges in stability prediction, since accurate phase stability diagrams rely on accurate calculation of formation enthalpies across a wide range of competing phases from metals to semiconductors to insulators. 
The challenge is complicated by the fact that there are many materials that are almost, but not, thermodynamically stable. 
That, combined with the uncertainty in DFT makes the materials search problem very challenging.  

Finally, Fig.~\ref{auinse2} shows the phase stability diagram for AuInSe$_2$ according to PBE and HSE06. 
Unlike the other two compounds, the line for In$_2$Se$_3$ line is not shown because it is located below the triangle. 
Thus, according to both PBE and HSE06 AuInSe$_2$ is always unstable with respect to the decomposition into In$_2$Se$_3$ and Au$_2$Se, and we predict that under all circumstances it can only be present as a non-equilibrium phase. 
Given that both functionals predict this and that the In$_2$Se$_3$ line lies so far below the triangle, we do not believe this is a case of a ``false negative", which would require that the error in the calculated formation enthalpies be much larger than observed for the other chalcopyrites. 
Therefore, synthesis of AuInSe$_2$ likely needs to proceed through non-equilibrium or potentially high pressure routes.

\section{Conclusions}
In conclusion, we have carried out (i) an in-depth, systematic  comparison of the structural parameters and electronic structure of the chalcopyrites CuInSe$_2$, AgInSe$_2$, and AuInSe$_2$ using screened exchange hybrid density functional theory, and (ii) a comparison of the phase stability predictions in this material set according to PBE and HSE06 in order to assess the degree of accuracy attainable. 
Consistent with experiment, we find that the band gap of AgInSe$_2$ is larger than that of CuInSe$_2$ due to a reduced $p$-$d$ repulsion in the valence bands. 
According to our HSE06 calculations, AuInSe$_2$ is predicted to have a small band gap $\approx$ 0.16 eV. 
Regarding phase stability, for the known compounds CuInSe$_2$ and AgInSe$_2$ the calculated stabilities indicate that the compounds are stable in a range of chemical potential phase space in good agreement with experimental results.
We further predict that AuInSe$_2$ is not stable, and can only be present as a non-equilibrium phase. 
A comparison of the phase stabilities suggests that for these chalcopyrites typical PBE errors in the calculation of formation enthalpies can be as large as 0.20 eV/atom. 
However the use of HSE06 is observed to reduce this by approximately a factor of three.  

\begin{acknowledgments}
The authors gratefully acknowledge the support of the International Institute for Carbon Neutral Energy Research (WPI-I$^2$CNER), sponsored by the World Premier International Research Center Initiative (WPI), MEXT, Japan. Computational resources were provided by (i) the Extreme Science and Engineering Discovery Environment (XSEDE) allocation DMR-140007, which is supported by National Science Foundation grant number ACI-1053575, and (ii) the Illinois Campus Computing Cluster. We are grateful to L. K. Wagner for comments and useful discussions. 
\end{acknowledgments}

\appendix*

\section{Density of states of binary compounds}

\begin{figure*}[h]
\includegraphics{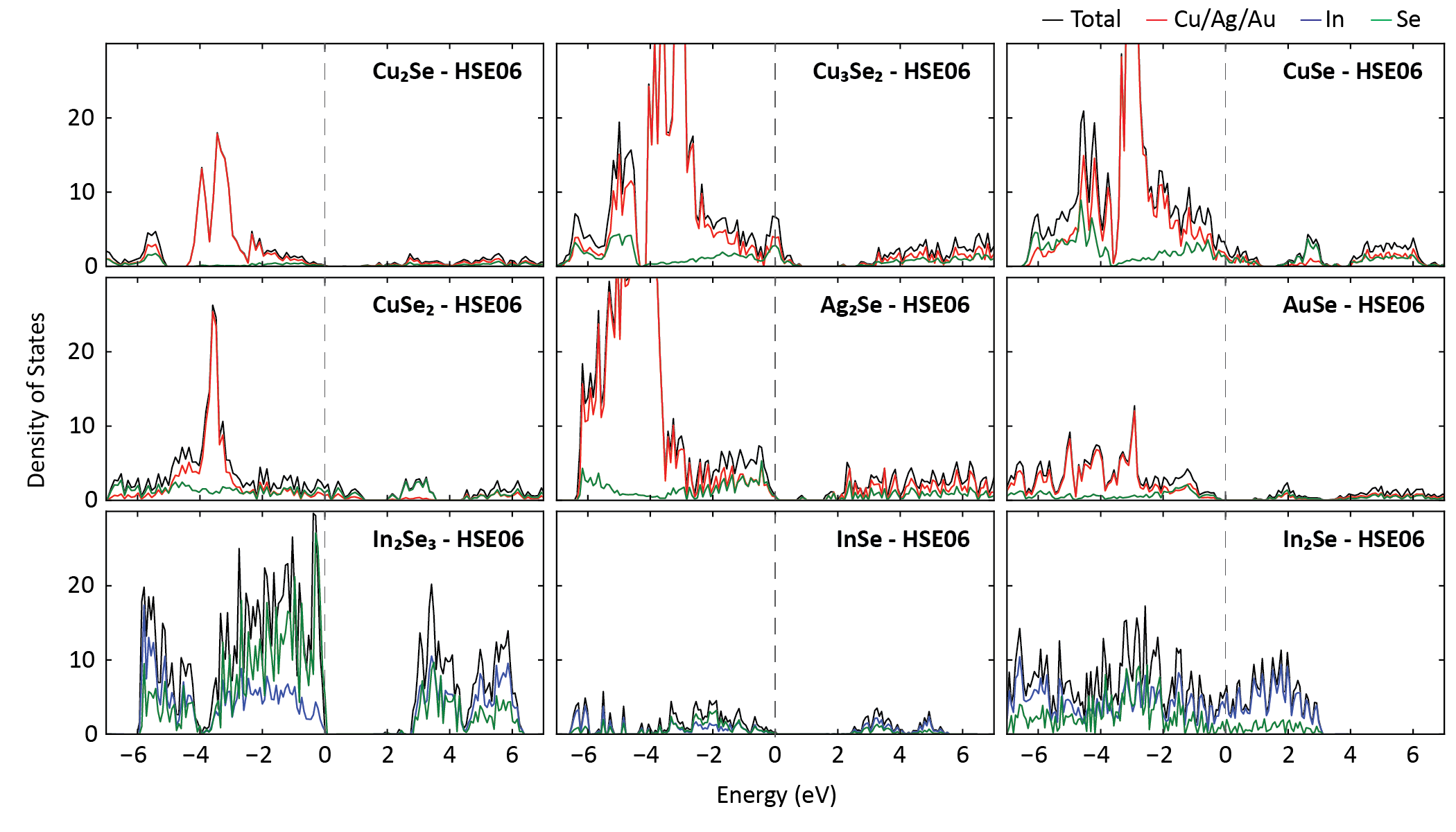}
\caption{\label{binary}
Density of states (DOS) for binary compounds calculated by DFT-HSE06. The black lines are total DOS, and red, blue and green lines are projected DOS of Cu/Ag/Au, In, and Se respectively. The Fermi level is shown by the dashed grey line.}
\end{figure*}

\bibliography{kim}

\begin{thebibliography}{44}%
\makeatletter
\providecommand \@ifxundefined [1]{%
 \@ifx{#1\undefined}
}%
\providecommand \@ifnum [1]{%
 \ifnum #1\expandafter \@firstoftwo
 \else \expandafter \@secondoftwo
 \fi
}%
\providecommand \@ifx [1]{%
 \ifx #1\expandafter \@firstoftwo
 \else \expandafter \@secondoftwo
 \fi
}%
\providecommand \natexlab [1]{#1}%
\providecommand \enquote  [1]{``#1''}%
\providecommand \bibnamefont  [1]{#1}%
\providecommand \bibfnamefont [1]{#1}%
\providecommand \citenamefont [1]{#1}%
\providecommand \href@noop [0]{\@secondoftwo}%
\providecommand \href [0]{\begingroup \@sanitize@url \@href}%
\providecommand \@href[1]{\@@startlink{#1}\@@href}%
\providecommand \@@href[1]{\endgroup#1\@@endlink}%
\providecommand \@sanitize@url [0]{\catcode `\\12\catcode `\$12\catcode
  `\&12\catcode `\#12\catcode `\^12\catcode `\_12\catcode `\%12\relax}%
\providecommand \@@startlink[1]{}%
\providecommand \@@endlink[0]{}%
\providecommand \url  [0]{\begingroup\@sanitize@url \@url }%
\providecommand \@url [1]{\endgroup\@href {#1}{\urlprefix }}%
\providecommand \urlprefix  [0]{URL }%
\providecommand \Eprint [0]{\href }%
\providecommand \doibase [0]{http://dx.doi.org/}%
\providecommand \selectlanguage [0]{\@gobble}%
\providecommand \bibinfo  [0]{\@secondoftwo}%
\providecommand \bibfield  [0]{\@secondoftwo}%
\providecommand \translation [1]{[#1]}%
\providecommand \BibitemOpen [0]{}%
\providecommand \bibitemStop [0]{}%
\providecommand \bibitemNoStop [0]{.\EOS\space}%
\providecommand \EOS [0]{\spacefactor3000\relax}%
\providecommand \BibitemShut  [1]{\csname bibitem#1\endcsname}%
\let\auto@bib@innerbib\@empty
\bibitem [{\citenamefont {Green}\ \emph {et~al.}(2016)\citenamefont {Green},
  \citenamefont {Emery}, \citenamefont {Hishikawa}, \citenamefont {Warta},\
  and\ \citenamefont {Dunlop}}]{PIP:PIP2728}%
  \BibitemOpen
  \bibfield  {author} {\bibinfo {author} {\bibfnamefont {M.~A.}\ \bibnamefont
  {Green}}, \bibinfo {author} {\bibfnamefont {K.}~\bibnamefont {Emery}},
  \bibinfo {author} {\bibfnamefont {Y.}~\bibnamefont {Hishikawa}}, \bibinfo
  {author} {\bibfnamefont {W.}~\bibnamefont {Warta}}, \ and\ \bibinfo {author}
  {\bibfnamefont {E.~D.}\ \bibnamefont {Dunlop}},\ }\href {\doibase
  10.1002/pip.2728} {\bibfield  {journal} {\bibinfo  {journal} {Progress in
  Photovoltaics: Research and Applications}\ }\textbf {\bibinfo {volume}
  {24}},\ \bibinfo {pages} {3} (\bibinfo {year} {2016})},\ \bibinfo {note}
  {pIP-15-272}\BibitemShut {NoStop}%
\bibitem [{\citenamefont {Pe\~{n}a Martin}\ \emph {et~al.}(2012)\citenamefont
  {Pe\~{n}a Martin}, \citenamefont {Rockett},\ and\ \citenamefont
  {Lyding}}]{:/content/avs/journal/jvsta/30/4/10.1116/1.4728160}%
  \BibitemOpen
  \bibfield  {author} {\bibinfo {author} {\bibfnamefont {P.}~\bibnamefont
  {Pe\~{n}a Martin}}, \bibinfo {author} {\bibfnamefont {A.~A.}\ \bibnamefont
  {Rockett}}, \ and\ \bibinfo {author} {\bibfnamefont {J.}~\bibnamefont
  {Lyding}},\ }\href {\doibase http://dx.doi.org/10.1116/1.4728160} {\bibfield
  {journal} {\bibinfo  {journal} {Journal of Vacuum Science \& Technology A}\
  }\textbf {\bibinfo {volume} {30}},\ \bibinfo {eid} {04D115} (\bibinfo {year}
  {2012}),\ http://dx.doi.org/10.1116/1.4728160}\BibitemShut {NoStop}%
\bibitem [{\citenamefont {Erslev}\ \emph {et~al.}(2011)\citenamefont {Erslev},
  \citenamefont {Lee}, \citenamefont {Hanket}, \citenamefont {Shafarman},\ and\
  \citenamefont {Cohen}}]{Erslev20117296}%
  \BibitemOpen
  \bibfield  {author} {\bibinfo {author} {\bibfnamefont {P.~T.}\ \bibnamefont
  {Erslev}}, \bibinfo {author} {\bibfnamefont {J.}~\bibnamefont {Lee}},
  \bibinfo {author} {\bibfnamefont {G.~M.}\ \bibnamefont {Hanket}}, \bibinfo
  {author} {\bibfnamefont {W.~N.}\ \bibnamefont {Shafarman}}, \ and\ \bibinfo
  {author} {\bibfnamefont {J.~D.}\ \bibnamefont {Cohen}},\ }\href {\doibase
  http://dx.doi.org/10.1016/j.tsf.2011.01.368} {\bibfield  {journal} {\bibinfo
  {journal} {Thin Solid Films}\ }\textbf {\bibinfo {volume} {519}},\ \bibinfo
  {pages} {7296 } (\bibinfo {year} {2011})},\ \bibinfo {note} {proceedings of
  the \{EMRS\} 2010 Spring Meeting Symposium M: Thin Film Chalcogenide
  Photovoltaic Materials}\BibitemShut {NoStop}%
\bibitem [{\citenamefont {Albornoz}\ \emph {et~al.}(2014)\citenamefont
  {Albornoz}, \citenamefont {L.}, \citenamefont {Merino},\ and\ \citenamefont
  {'{o}n}}]{Albornoz20141}%
  \BibitemOpen
  \bibfield  {author} {\bibinfo {author} {\bibfnamefont {J.}~\bibnamefont
  {Albornoz}}, \bibinfo {author} {\bibfnamefont {R.~R.}\ \bibnamefont {L.}},
  \bibinfo {author} {\bibfnamefont {J.}~\bibnamefont {Merino}}, \ and\ \bibinfo
  {author} {\bibfnamefont {M.~L.}\ \bibnamefont {'{o}n}},\ }\href {\doibase
  http://dx.doi.org/10.1016/j.jpcs.2013.08.003} {\bibfield  {journal} {\bibinfo
   {journal} {Journal of Physics and Chemistry of Solids}\ }\textbf {\bibinfo
  {volume} {75}},\ \bibinfo {pages} {1 } (\bibinfo {year} {2014})}\BibitemShut
  {NoStop}%
\bibitem [{\citenamefont {Feng}\ \emph {et~al.}(2011)\citenamefont {Feng},
  \citenamefont {Xiao}, \citenamefont {Ding},\ and\ \citenamefont
  {Yao}}]{PhysRevLett.106.016402}%
  \BibitemOpen
  \bibfield  {author} {\bibinfo {author} {\bibfnamefont {W.}~\bibnamefont
  {Feng}}, \bibinfo {author} {\bibfnamefont {D.}~\bibnamefont {Xiao}}, \bibinfo
  {author} {\bibfnamefont {J.}~\bibnamefont {Ding}}, \ and\ \bibinfo {author}
  {\bibfnamefont {Y.}~\bibnamefont {Yao}},\ }\href {\doibase
  10.1103/PhysRevLett.106.016402} {\bibfield  {journal} {\bibinfo  {journal}
  {Phys. Rev. Lett.}\ }\textbf {\bibinfo {volume} {106}},\ \bibinfo {pages}
  {016402} (\bibinfo {year} {2011})}\BibitemShut {NoStop}%
\bibitem [{\citenamefont {Feng}\ and\ \citenamefont {Yao}(2012)}]{Feng2012}%
  \BibitemOpen
  \bibfield  {author} {\bibinfo {author} {\bibfnamefont {W.}~\bibnamefont
  {Feng}}\ and\ \bibinfo {author} {\bibfnamefont {Y.}~\bibnamefont {Yao}},\
  }\href {\doibase 10.1007/s11433-012-4929-9} {\bibfield  {journal} {\bibinfo
  {journal} {Science China Physics, Mechanics and Astronomy}\ }\textbf
  {\bibinfo {volume} {55}},\ \bibinfo {pages} {2199} (\bibinfo {year}
  {2012})}\BibitemShut {NoStop}%
\bibitem [{\citenamefont {Di~Giulio}\ \emph {et~al.}(1983)\citenamefont
  {Di~Giulio}, \citenamefont {Micocci}, \citenamefont {Rizzo},\ and\
  \citenamefont {Tepore}}]{:/content/aip/journal/jap/54/10/10.1063/1.331808}%
  \BibitemOpen
  \bibfield  {author} {\bibinfo {author} {\bibfnamefont {M.}~\bibnamefont
  {Di~Giulio}}, \bibinfo {author} {\bibfnamefont {G.}~\bibnamefont {Micocci}},
  \bibinfo {author} {\bibfnamefont {A.}~\bibnamefont {Rizzo}}, \ and\ \bibinfo
  {author} {\bibfnamefont {A.}~\bibnamefont {Tepore}},\ }\href {\doibase
  http://dx.doi.org/10.1063/1.331808} {\bibfield  {journal} {\bibinfo
  {journal} {Journal of Applied Physics}\ }\textbf {\bibinfo {volume} {54}},\
  \bibinfo {pages} {5839} (\bibinfo {year} {1983})}\BibitemShut {NoStop}%
\bibitem [{\citenamefont {Mart\'{i}nez?Pastor}\ \emph
  {et~al.}(1987)\citenamefont {Mart\'{i}nez?Pastor}, \citenamefont {Segura},
  \citenamefont {Vald\'{e}s},\ and\ \citenamefont
  {Chevy}}]{:/content/aip/journal/jap/62/4/10.1063/1.339627}%
  \BibitemOpen
  \bibfield  {author} {\bibinfo {author} {\bibfnamefont {J.}~\bibnamefont
  {Mart\'{i}nez?Pastor}}, \bibinfo {author} {\bibfnamefont {A.}~\bibnamefont
  {Segura}}, \bibinfo {author} {\bibfnamefont {J.~L.}\ \bibnamefont
  {Vald\'{e}s}}, \ and\ \bibinfo {author} {\bibfnamefont {A.}~\bibnamefont
  {Chevy}},\ }\href {\doibase http://dx.doi.org/10.1063/1.339627} {\bibfield
  {journal} {\bibinfo  {journal} {Journal of Applied Physics}\ }\textbf
  {\bibinfo {volume} {62}},\ \bibinfo {pages} {1477} (\bibinfo {year}
  {1987})}\BibitemShut {NoStop}%
\bibitem [{\citenamefont {Sun}\ \emph {et~al.}(2006)\citenamefont {Sun},
  \citenamefont {Yu}, \citenamefont {Ng}, \citenamefont {Nguyen},\ and\
  \citenamefont
  {Meyyappan}}]{:/content/aip/journal/apl/89/23/10.1063/1.2388890}%
  \BibitemOpen
  \bibfield  {author} {\bibinfo {author} {\bibfnamefont {X.}~\bibnamefont
  {Sun}}, \bibinfo {author} {\bibfnamefont {B.}~\bibnamefont {Yu}}, \bibinfo
  {author} {\bibfnamefont {G.}~\bibnamefont {Ng}}, \bibinfo {author}
  {\bibfnamefont {T.~D.}\ \bibnamefont {Nguyen}}, \ and\ \bibinfo {author}
  {\bibfnamefont {M.}~\bibnamefont {Meyyappan}},\ }\href {\doibase
  http://dx.doi.org/10.1063/1.2388890} {\bibfield  {journal} {\bibinfo
  {journal} {Applied Physics Letters}\ }\textbf {\bibinfo {volume} {89}},\
  \bibinfo {eid} {233121} (\bibinfo {year} {2006}),\
  http://dx.doi.org/10.1063/1.2388890}\BibitemShut {NoStop}%
\bibitem [{\citenamefont {Pohl}\ and\ \citenamefont
  {Albe}(2010)}]{:/content/aip/journal/jap/108/2/10.1063/1.3456161}%
  \BibitemOpen
  \bibfield  {author} {\bibinfo {author} {\bibfnamefont {J.}~\bibnamefont
  {Pohl}}\ and\ \bibinfo {author} {\bibfnamefont {K.}~\bibnamefont {Albe}},\
  }\href {\doibase http://dx.doi.org/10.1063/1.3456161} {\bibfield  {journal}
  {\bibinfo  {journal} {Journal of Applied Physics}\ }\textbf {\bibinfo
  {volume} {108}},\ \bibinfo {eid} {023509} (\bibinfo {year}
  {2010})}\BibitemShut {NoStop}%
\bibitem [{\citenamefont {Pohl}\ and\ \citenamefont
  {Albe}(2013)}]{PhysRevB.87.245203}%
  \BibitemOpen
  \bibfield  {author} {\bibinfo {author} {\bibfnamefont {J.}~\bibnamefont
  {Pohl}}\ and\ \bibinfo {author} {\bibfnamefont {K.}~\bibnamefont {Albe}},\
  }\href {\doibase 10.1103/PhysRevB.87.245203} {\bibfield  {journal} {\bibinfo
  {journal} {Phys. Rev. B}\ }\textbf {\bibinfo {volume} {87}},\ \bibinfo
  {pages} {245203} (\bibinfo {year} {2013})}\BibitemShut {NoStop}%
\bibitem [{\citenamefont {Vidal}\ \emph {et~al.}(2010)\citenamefont {Vidal},
  \citenamefont {Botti}, \citenamefont {Olsson}, \citenamefont {Guillemoles},\
  and\ \citenamefont {Reining}}]{PhysRevLett.104.056401}%
  \BibitemOpen
  \bibfield  {author} {\bibinfo {author} {\bibfnamefont {J.}~\bibnamefont
  {Vidal}}, \bibinfo {author} {\bibfnamefont {S.}~\bibnamefont {Botti}},
  \bibinfo {author} {\bibfnamefont {P.}~\bibnamefont {Olsson}}, \bibinfo
  {author} {\bibfnamefont {J.-F. m.~c.}\ \bibnamefont {Guillemoles}}, \ and\
  \bibinfo {author} {\bibfnamefont {L.}~\bibnamefont {Reining}},\ }\href
  {\doibase 10.1103/PhysRevLett.104.056401} {\bibfield  {journal} {\bibinfo
  {journal} {Phys. Rev. Lett.}\ }\textbf {\bibinfo {volume} {104}},\ \bibinfo
  {pages} {056401} (\bibinfo {year} {2010})}\BibitemShut {NoStop}%
\bibitem [{\citenamefont {Sharma}\ \emph {et~al.}(2014)\citenamefont {Sharma},
  \citenamefont {Verma},\ and\ \citenamefont {Jindal}}]{Sharma201497}%
  \BibitemOpen
  \bibfield  {author} {\bibinfo {author} {\bibfnamefont {S.}~\bibnamefont
  {Sharma}}, \bibinfo {author} {\bibfnamefont {A.}~\bibnamefont {Verma}}, \
  and\ \bibinfo {author} {\bibfnamefont {V.}~\bibnamefont {Jindal}},\ }\href
  {\doibase http://dx.doi.org/10.1016/j.physb.2013.12.031} {\bibfield
  {journal} {\bibinfo  {journal} {Physica B: Condensed Matter}\ }\textbf
  {\bibinfo {volume} {438}},\ \bibinfo {pages} {97 } (\bibinfo {year}
  {2014})}\BibitemShut {NoStop}%
\bibitem [{\citenamefont {Fan}\ \emph {et~al.}(2014)\citenamefont {Fan},
  \citenamefont {Wu},\ and\ \citenamefont {Yu}}]{C3EE41437J}%
  \BibitemOpen
  \bibfield  {author} {\bibinfo {author} {\bibfnamefont {F.-J.}\ \bibnamefont
  {Fan}}, \bibinfo {author} {\bibfnamefont {L.}~\bibnamefont {Wu}}, \ and\
  \bibinfo {author} {\bibfnamefont {S.-H.}\ \bibnamefont {Yu}},\ }\href
  {\doibase 10.1039/C3EE41437J} {\bibfield  {journal} {\bibinfo  {journal}
  {Energy Environ. Sci.}\ }\textbf {\bibinfo {volume} {7}},\ \bibinfo {pages}
  {190} (\bibinfo {year} {2014})}\BibitemShut {NoStop}%
\bibitem [{\citenamefont {Jaffe}\ and\ \citenamefont
  {Zunger}(1983)}]{PhysRevB.27.5176}%
  \BibitemOpen
  \bibfield  {author} {\bibinfo {author} {\bibfnamefont {J.~E.}\ \bibnamefont
  {Jaffe}}\ and\ \bibinfo {author} {\bibfnamefont {A.}~\bibnamefont {Zunger}},\
  }\href {\doibase 10.1103/PhysRevB.27.5176} {\bibfield  {journal} {\bibinfo
  {journal} {Phys. Rev. B}\ }\textbf {\bibinfo {volume} {27}},\ \bibinfo
  {pages} {5176} (\bibinfo {year} {1983})}\BibitemShut {NoStop}%
\bibitem [{\citenamefont {Jaffe}\ and\ \citenamefont
  {Zunger}(1984)}]{PhysRevB.29.1882}%
  \BibitemOpen
  \bibfield  {author} {\bibinfo {author} {\bibfnamefont {J.~E.}\ \bibnamefont
  {Jaffe}}\ and\ \bibinfo {author} {\bibfnamefont {A.}~\bibnamefont {Zunger}},\
  }\href {\doibase 10.1103/PhysRevB.29.1882} {\bibfield  {journal} {\bibinfo
  {journal} {Phys. Rev. B}\ }\textbf {\bibinfo {volume} {29}},\ \bibinfo
  {pages} {1882} (\bibinfo {year} {1984})}\BibitemShut {NoStop}%
\bibitem [{\citenamefont {Heyd}\ \emph {et~al.}(2003)\citenamefont {Heyd},
  \citenamefont {Scuseria},\ and\ \citenamefont
  {Ernzerhof}}]{:/content/aip/journal/jcp/118/18/10.1063/1.1564060}%
  \BibitemOpen
  \bibfield  {author} {\bibinfo {author} {\bibfnamefont {J.}~\bibnamefont
  {Heyd}}, \bibinfo {author} {\bibfnamefont {G.~E.}\ \bibnamefont {Scuseria}},
  \ and\ \bibinfo {author} {\bibfnamefont {M.}~\bibnamefont {Ernzerhof}},\
  }\href {\doibase http://dx.doi.org/10.1063/1.1564060} {\bibfield  {journal}
  {\bibinfo  {journal} {The Journal of Chemical Physics}\ }\textbf {\bibinfo
  {volume} {118}},\ \bibinfo {pages} {8207} (\bibinfo {year}
  {2003})}\BibitemShut {NoStop}%
\bibitem [{\citenamefont {Heyd}\ and\ \citenamefont
  {Scuseria}(2004)}]{:/content/aip/journal/jcp/121/3/10.1063/1.1760074}%
  \BibitemOpen
  \bibfield  {author} {\bibinfo {author} {\bibfnamefont {J.}~\bibnamefont
  {Heyd}}\ and\ \bibinfo {author} {\bibfnamefont {G.~E.}\ \bibnamefont
  {Scuseria}},\ }\href {\doibase http://dx.doi.org/10.1063/1.1760074}
  {\bibfield  {journal} {\bibinfo  {journal} {The Journal of Chemical Physics}\
  }\textbf {\bibinfo {volume} {121}},\ \bibinfo {pages} {1187} (\bibinfo {year}
  {2004})}\BibitemShut {NoStop}%
\bibitem [{\citenamefont {Krukau}\ \emph {et~al.}(2006)\citenamefont {Krukau},
  \citenamefont {Vydrov}, \citenamefont {Izmaylov},\ and\ \citenamefont
  {Scuseria}}]{:/content/aip/journal/jcp/125/22/10.1063/1.2404663}%
  \BibitemOpen
  \bibfield  {author} {\bibinfo {author} {\bibfnamefont {A.~V.}\ \bibnamefont
  {Krukau}}, \bibinfo {author} {\bibfnamefont {O.~A.}\ \bibnamefont {Vydrov}},
  \bibinfo {author} {\bibfnamefont {A.~F.}\ \bibnamefont {Izmaylov}}, \ and\
  \bibinfo {author} {\bibfnamefont {G.~E.}\ \bibnamefont {Scuseria}},\ }\href
  {\doibase http://dx.doi.org/10.1063/1.2404663} {\bibfield  {journal}
  {\bibinfo  {journal} {The Journal of Chemical Physics}\ }\textbf {\bibinfo
  {volume} {125}},\ \bibinfo {eid} {224106} (\bibinfo {year}
  {2006})}\BibitemShut {NoStop}%
\bibitem [{\citenamefont {Perdew}\ \emph {et~al.}(1996)\citenamefont {Perdew},
  \citenamefont {Burke},\ and\ \citenamefont
  {Ernzerhof}}]{PhysRevLett.77.3865}%
  \BibitemOpen
  \bibfield  {author} {\bibinfo {author} {\bibfnamefont {J.~P.}\ \bibnamefont
  {Perdew}}, \bibinfo {author} {\bibfnamefont {K.}~\bibnamefont {Burke}}, \
  and\ \bibinfo {author} {\bibfnamefont {M.}~\bibnamefont {Ernzerhof}},\ }\href
  {\doibase 10.1103/PhysRevLett.77.3865} {\bibfield  {journal} {\bibinfo
  {journal} {Phys. Rev. Lett.}\ }\textbf {\bibinfo {volume} {77}},\ \bibinfo
  {pages} {3865} (\bibinfo {year} {1996})}\BibitemShut {NoStop}%
\bibitem [{\citenamefont {Dey}\ \emph {et~al.}(2014)\citenamefont {Dey},
  \citenamefont {Bible}, \citenamefont {Datta}, \citenamefont {Broderick},
  \citenamefont {Jasinski}, \citenamefont {Sunkara}, \citenamefont {Menon},\
  and\ \citenamefont {Rajan}}]{Dey2014185}%
  \BibitemOpen
  \bibfield  {author} {\bibinfo {author} {\bibfnamefont {P.}~\bibnamefont
  {Dey}}, \bibinfo {author} {\bibfnamefont {J.}~\bibnamefont {Bible}}, \bibinfo
  {author} {\bibfnamefont {S.}~\bibnamefont {Datta}}, \bibinfo {author}
  {\bibfnamefont {S.}~\bibnamefont {Broderick}}, \bibinfo {author}
  {\bibfnamefont {J.}~\bibnamefont {Jasinski}}, \bibinfo {author}
  {\bibfnamefont {M.}~\bibnamefont {Sunkara}}, \bibinfo {author} {\bibfnamefont
  {M.}~\bibnamefont {Menon}}, \ and\ \bibinfo {author} {\bibfnamefont
  {K.}~\bibnamefont {Rajan}},\ }\href {\doibase
  http://dx.doi.org/10.1016/j.commatsci.2013.10.016} {\bibfield  {journal}
  {\bibinfo  {journal} {Computational Materials Science}\ }\textbf {\bibinfo
  {volume} {83}},\ \bibinfo {pages} {185 } (\bibinfo {year}
  {2014})}\BibitemShut {NoStop}%
\bibitem [{\citenamefont {Hohenberg}\ and\ \citenamefont
  {Kohn}(1964)}]{PhysRev.136.B864}%
  \BibitemOpen
  \bibfield  {author} {\bibinfo {author} {\bibfnamefont {P.}~\bibnamefont
  {Hohenberg}}\ and\ \bibinfo {author} {\bibfnamefont {W.}~\bibnamefont
  {Kohn}},\ }\href {\doibase 10.1103/PhysRev.136.B864} {\bibfield  {journal}
  {\bibinfo  {journal} {Phys. Rev.}\ }\textbf {\bibinfo {volume} {136}},\
  \bibinfo {pages} {B864} (\bibinfo {year} {1964})}\BibitemShut {NoStop}%
\bibitem [{\citenamefont {Kohn}\ and\ \citenamefont
  {Sham}(1965)}]{PhysRev.140.A1133}%
  \BibitemOpen
  \bibfield  {author} {\bibinfo {author} {\bibfnamefont {W.}~\bibnamefont
  {Kohn}}\ and\ \bibinfo {author} {\bibfnamefont {L.~J.}\ \bibnamefont
  {Sham}},\ }\href {\doibase 10.1103/PhysRev.140.A1133} {\bibfield  {journal}
  {\bibinfo  {journal} {Phys. Rev.}\ }\textbf {\bibinfo {volume} {140}},\
  \bibinfo {pages} {A1133} (\bibinfo {year} {1965})}\BibitemShut {NoStop}%
\bibitem [{\citenamefont {Kresse}\ and\ \citenamefont
  {Furthm\"uller}(1996{\natexlab{a}})}]{Kresse199615}%
  \BibitemOpen
  \bibfield  {author} {\bibinfo {author} {\bibfnamefont {G.}~\bibnamefont
  {Kresse}}\ and\ \bibinfo {author} {\bibfnamefont {J.}~\bibnamefont
  {Furthm\"uller}},\ }\href {\doibase
  http://dx.doi.org/10.1016/0927-0256(96)00008-0} {\bibfield  {journal}
  {\bibinfo  {journal} {Computational Materials Science}\ }\textbf {\bibinfo
  {volume} {6}},\ \bibinfo {pages} {15 } (\bibinfo {year}
  {1996}{\natexlab{a}})}\BibitemShut {NoStop}%
\bibitem [{\citenamefont {Kresse}\ and\ \citenamefont
  {Furthm\"uller}(1996{\natexlab{b}})}]{PhysRevB.54.11169}%
  \BibitemOpen
  \bibfield  {author} {\bibinfo {author} {\bibfnamefont {G.}~\bibnamefont
  {Kresse}}\ and\ \bibinfo {author} {\bibfnamefont {J.}~\bibnamefont
  {Furthm\"uller}},\ }\href {\doibase 10.1103/PhysRevB.54.11169} {\bibfield
  {journal} {\bibinfo  {journal} {Phys. Rev. B}\ }\textbf {\bibinfo {volume}
  {54}},\ \bibinfo {pages} {11169} (\bibinfo {year}
  {1996}{\natexlab{b}})}\BibitemShut {NoStop}%
\bibitem [{\citenamefont {Bl\"ochl}(1994)}]{PhysRevB.50.17953}%
  \BibitemOpen
  \bibfield  {author} {\bibinfo {author} {\bibfnamefont {P.~E.}\ \bibnamefont
  {Bl\"ochl}},\ }\href {\doibase 10.1103/PhysRevB.50.17953} {\bibfield
  {journal} {\bibinfo  {journal} {Phys. Rev. B}\ }\textbf {\bibinfo {volume}
  {50}},\ \bibinfo {pages} {17953} (\bibinfo {year} {1994})}\BibitemShut
  {NoStop}%
\bibitem [{\citenamefont {Kresse}\ and\ \citenamefont
  {Joubert}(1999)}]{PhysRevB.59.1758}%
  \BibitemOpen
  \bibfield  {author} {\bibinfo {author} {\bibfnamefont {G.}~\bibnamefont
  {Kresse}}\ and\ \bibinfo {author} {\bibfnamefont {D.}~\bibnamefont
  {Joubert}},\ }\href {\doibase 10.1103/PhysRevB.59.1758} {\bibfield  {journal}
  {\bibinfo  {journal} {Phys. Rev. B}\ }\textbf {\bibinfo {volume} {59}},\
  \bibinfo {pages} {1758} (\bibinfo {year} {1999})}\BibitemShut {NoStop}%
\bibitem [{\citenamefont {Oikkonen}\ \emph {et~al.}(2011)\citenamefont
  {Oikkonen}, \citenamefont {Ganchenkova}, \citenamefont {Seitsonen},\ and\
  \citenamefont {Nieminen}}]{0953-8984-23-42-422202}%
  \BibitemOpen
  \bibfield  {author} {\bibinfo {author} {\bibfnamefont {L.~E.}\ \bibnamefont
  {Oikkonen}}, \bibinfo {author} {\bibfnamefont {M.~G.}\ \bibnamefont
  {Ganchenkova}}, \bibinfo {author} {\bibfnamefont {A.~P.}\ \bibnamefont
  {Seitsonen}}, \ and\ \bibinfo {author} {\bibfnamefont {R.~M.}\ \bibnamefont
  {Nieminen}},\ }\href {http://stacks.iop.org/0953-8984/23/i=42/a=422202}
  {\bibfield  {journal} {\bibinfo  {journal} {Journal of Physics: Condensed
  Matter}\ }\textbf {\bibinfo {volume} {23}},\ \bibinfo {pages} {422202}
  (\bibinfo {year} {2011})}\BibitemShut {NoStop}%
\bibitem [{\citenamefont {Momma}\ and\ \citenamefont
  {Izumi}(2011)}]{Momma:db5098}%
  \BibitemOpen
  \bibfield  {author} {\bibinfo {author} {\bibfnamefont {K.}~\bibnamefont
  {Momma}}\ and\ \bibinfo {author} {\bibfnamefont {F.}~\bibnamefont {Izumi}},\
  }\href {\doibase 10.1107/S0021889811038970} {\bibfield  {journal} {\bibinfo
  {journal} {Journal of Applied Crystallography}\ }\textbf {\bibinfo {volume}
  {44}},\ \bibinfo {pages} {1272} (\bibinfo {year} {2011})}\BibitemShut
  {NoStop}%
\bibitem [{\citenamefont {Knight}(1992)}]{KNIGHT1992161}%
  \BibitemOpen
  \bibfield  {author} {\bibinfo {author} {\bibfnamefont {K.}~\bibnamefont
  {Knight}},\ }\href {\doibase http://dx.doi.org/10.1016/0025-5408(92)90209-I}
  {\bibfield  {journal} {\bibinfo  {journal} {Materials Research Bulletin}\
  }\textbf {\bibinfo {volume} {27}},\ \bibinfo {pages} {161 } (\bibinfo {year}
  {1992})}\BibitemShut {NoStop}%
\bibitem [{\citenamefont {Benoit}\ \emph {et~al.}(1980)\citenamefont {Benoit},
  \citenamefont {Charpin}, \citenamefont {Lesueur},\ and\ \citenamefont
  {Djega-Mariadassou}}]{1347-4065-19-S3-85}%
  \BibitemOpen
  \bibfield  {author} {\bibinfo {author} {\bibfnamefont {P.}~\bibnamefont
  {Benoit}}, \bibinfo {author} {\bibfnamefont {P.}~\bibnamefont {Charpin}},
  \bibinfo {author} {\bibfnamefont {R.}~\bibnamefont {Lesueur}}, \ and\
  \bibinfo {author} {\bibfnamefont {C.}~\bibnamefont {Djega-Mariadassou}},\
  }\href {http://stacks.iop.org/1347-4065/19/i=S3/a=85} {\bibfield  {journal}
  {\bibinfo  {journal} {Japanese Journal of Applied Physics}\ }\textbf
  {\bibinfo {volume} {19}},\ \bibinfo {pages} {85} (\bibinfo {year}
  {1980})}\BibitemShut {NoStop}%
\bibitem [{\citenamefont {SHAY}\ and\ \citenamefont
  {WERNICK}(1975)}]{SHAY1975110}%
  \BibitemOpen
  \bibfield  {author} {\bibinfo {author} {\bibfnamefont {J.}~\bibnamefont
  {SHAY}}\ and\ \bibinfo {author} {\bibfnamefont {J.}~\bibnamefont {WERNICK}},\
  }in\ \href {\doibase http://dx.doi.org/10.1016/B978-0-08-017883-7.50009-3}
  {\emph {\bibinfo {booktitle} {Ternary Chalcopyrite Semiconductors: Growth,
  Electronic Properties, and Applications}}},\ \bibinfo {series} {International
  Series in the Science of the Solid State}, Vol.~\bibinfo {volume} {7},\
  \bibinfo {editor} {edited by\ \bibinfo {editor} {\bibfnamefont {J.~S.}\
  \bibnamefont {WERNICK}}}\ (\bibinfo  {publisher} {Pergamon},\ \bibinfo {year}
  {1975})\ pp.\ \bibinfo {pages} {110 -- 128}\BibitemShut {NoStop}%
\bibitem [{\citenamefont {Hofmann}\ and\ \citenamefont
  {Pettenkofer}(2011)}]{PhysRevB.84.115109}%
  \BibitemOpen
  \bibfield  {author} {\bibinfo {author} {\bibfnamefont {A.}~\bibnamefont
  {Hofmann}}\ and\ \bibinfo {author} {\bibfnamefont {C.}~\bibnamefont
  {Pettenkofer}},\ }\href {\doibase 10.1103/PhysRevB.84.115109} {\bibfield
  {journal} {\bibinfo  {journal} {Phys. Rev. B}\ }\textbf {\bibinfo {volume}
  {84}},\ \bibinfo {pages} {115109} (\bibinfo {year} {2011})}\BibitemShut
  {NoStop}%
\bibitem [{\citenamefont {Zhang}\ \emph {et~al.}(2013)\citenamefont {Zhang},
  \citenamefont {Zhang}, \citenamefont {Gao}, \citenamefont {Abtew},
  \citenamefont {Wang}, \citenamefont {Zhang},\ and\ \citenamefont
  {Zhang}}]{inversion}%
  \BibitemOpen
  \bibfield  {author} {\bibinfo {author} {\bibfnamefont {Y.}~\bibnamefont
  {Zhang}}, \bibinfo {author} {\bibfnamefont {J.}~\bibnamefont {Zhang}},
  \bibinfo {author} {\bibfnamefont {W.}~\bibnamefont {Gao}}, \bibinfo {author}
  {\bibfnamefont {T.~A.}\ \bibnamefont {Abtew}}, \bibinfo {author}
  {\bibfnamefont {Y.}~\bibnamefont {Wang}}, \bibinfo {author} {\bibfnamefont
  {P.}~\bibnamefont {Zhang}}, \ and\ \bibinfo {author} {\bibfnamefont
  {W.}~\bibnamefont {Zhang}},\ }\href {\doibase
  http://dx.doi.org/10.1063/1.4828864} {\bibfield  {journal} {\bibinfo
  {journal} {The Journal of Chemical Physics}\ }\textbf {\bibinfo {volume}
  {139}},\ \bibinfo {eid} {184706} (\bibinfo {year} {2013}),\
  http://dx.doi.org/10.1063/1.4828864}\BibitemShut {NoStop}%
\bibitem [{\citenamefont {Lisensky}\ \emph {et~al.}(1992)\citenamefont
  {Lisensky}, \citenamefont {Penn}, \citenamefont {Geselbracht},\ and\
  \citenamefont {Ellis}}]{doi:10.1021/ed069p151}%
  \BibitemOpen
  \bibfield  {author} {\bibinfo {author} {\bibfnamefont {G.~C.}\ \bibnamefont
  {Lisensky}}, \bibinfo {author} {\bibfnamefont {R.}~\bibnamefont {Penn}},
  \bibinfo {author} {\bibfnamefont {M.~J.}\ \bibnamefont {Geselbracht}}, \ and\
  \bibinfo {author} {\bibfnamefont {A.~B.}\ \bibnamefont {Ellis}},\ }\href
  {\doibase 10.1021/ed069p151} {\bibfield  {journal} {\bibinfo  {journal}
  {Journal of Chemical Education}\ }\textbf {\bibinfo {volume} {69}},\ \bibinfo
  {pages} {151} (\bibinfo {year} {1992})},\ \Eprint
  {http://arxiv.org/abs/http://dx.doi.org/10.1021/ed069p151}
  {http://dx.doi.org/10.1021/ed069p151} \BibitemShut {NoStop}%
\bibitem [{\citenamefont {Asta}(2014)}]{Asta2014}%
  \BibitemOpen
  \bibfield  {author} {\bibinfo {author} {\bibfnamefont {M.}~\bibnamefont
  {Asta}},\ }\href {\doibase 10.1007/s11837-014-0887-1} {\bibfield  {journal}
  {\bibinfo  {journal} {JOM}\ }\textbf {\bibinfo {volume} {66}},\ \bibinfo
  {pages} {364} (\bibinfo {year} {2014})}\BibitemShut {NoStop}%
\bibitem [{\citenamefont {Jain}\ \emph {et~al.}(2013)\citenamefont {Jain},
  \citenamefont {Ong}, \citenamefont {Hautier}, \citenamefont {Chen},
  \citenamefont {Richards}, \citenamefont {Dacek}, \citenamefont {Cholia},
  \citenamefont {Gunter}, \citenamefont {Skinner}, \citenamefont {Ceder},\ and\
  \citenamefont {Persson}}]{Jain2013}%
  \BibitemOpen
  \bibfield  {author} {\bibinfo {author} {\bibfnamefont {A.}~\bibnamefont
  {Jain}}, \bibinfo {author} {\bibfnamefont {S.~P.}\ \bibnamefont {Ong}},
  \bibinfo {author} {\bibfnamefont {G.}~\bibnamefont {Hautier}}, \bibinfo
  {author} {\bibfnamefont {W.}~\bibnamefont {Chen}}, \bibinfo {author}
  {\bibfnamefont {W.~D.}\ \bibnamefont {Richards}}, \bibinfo {author}
  {\bibfnamefont {S.}~\bibnamefont {Dacek}}, \bibinfo {author} {\bibfnamefont
  {S.}~\bibnamefont {Cholia}}, \bibinfo {author} {\bibfnamefont
  {D.}~\bibnamefont {Gunter}}, \bibinfo {author} {\bibfnamefont
  {D.}~\bibnamefont {Skinner}}, \bibinfo {author} {\bibfnamefont
  {G.}~\bibnamefont {Ceder}}, \ and\ \bibinfo {author} {\bibfnamefont {K.~a.}\
  \bibnamefont {Persson}},\ }\href {\doibase 10.1063/1.4812323} {\bibfield
  {journal} {\bibinfo  {journal} {APL Materials}\ }\textbf {\bibinfo {volume}
  {1}},\ \bibinfo {pages} {011002} (\bibinfo {year} {2013})}\BibitemShut
  {NoStop}%
\bibitem [{\citenamefont {Stevanovi\ifmmode~\acute{c}\else \'{c}\fi{}}\ \emph
  {et~al.}(2012)\citenamefont {Stevanovi\ifmmode~\acute{c}\else \'{c}\fi{}},
  \citenamefont {Lany}, \citenamefont {Zhang},\ and\ \citenamefont
  {Zunger}}]{PhysRevB.85.115104}%
  \BibitemOpen
  \bibfield  {author} {\bibinfo {author} {\bibfnamefont {V.}~\bibnamefont
  {Stevanovi\ifmmode~\acute{c}\else \'{c}\fi{}}}, \bibinfo {author}
  {\bibfnamefont {S.}~\bibnamefont {Lany}}, \bibinfo {author} {\bibfnamefont
  {X.}~\bibnamefont {Zhang}}, \ and\ \bibinfo {author} {\bibfnamefont
  {A.}~\bibnamefont {Zunger}},\ }\href {\doibase 10.1103/PhysRevB.85.115104}
  {\bibfield  {journal} {\bibinfo  {journal} {Phys. Rev. B}\ }\textbf {\bibinfo
  {volume} {85}},\ \bibinfo {pages} {115104} (\bibinfo {year}
  {2012})}\BibitemShut {NoStop}%
\bibitem [{\citenamefont {Saal}\ \emph {et~al.}(2013)\citenamefont {Saal},
  \citenamefont {Kirklin}, \citenamefont {Aykol}, \citenamefont {Meredig},\
  and\ \citenamefont {Wolverton}}]{Saal2013}%
  \BibitemOpen
  \bibfield  {author} {\bibinfo {author} {\bibfnamefont {J.~E.}\ \bibnamefont
  {Saal}}, \bibinfo {author} {\bibfnamefont {S.}~\bibnamefont {Kirklin}},
  \bibinfo {author} {\bibfnamefont {M.}~\bibnamefont {Aykol}}, \bibinfo
  {author} {\bibfnamefont {B.}~\bibnamefont {Meredig}}, \ and\ \bibinfo
  {author} {\bibfnamefont {C.}~\bibnamefont {Wolverton}},\ }\href {\doibase
  10.1007/s11837-013-0755-4} {\bibfield  {journal} {\bibinfo  {journal} {JOM}\
  }\textbf {\bibinfo {volume} {65}},\ \bibinfo {pages} {1501} (\bibinfo {year}
  {2013})}\BibitemShut {NoStop}%
\bibitem [{\citenamefont {Berger}\ \emph {et~al.}(1973)\citenamefont {Berger},
  \citenamefont {Bondar}, \citenamefont {Lebedev}, \citenamefont {Molodyk},\
  and\ \citenamefont {Strelchenko}}]{formE_CIS_AIS}%
  \BibitemOpen
  \bibfield  {author} {\bibinfo {author} {\bibfnamefont {L.~I.}\ \bibnamefont
  {Berger}}, \bibinfo {author} {\bibfnamefont {S.}~\bibnamefont {Bondar}},
  \bibinfo {author} {\bibfnamefont {V.~V.}\ \bibnamefont {Lebedev}}, \bibinfo
  {author} {\bibfnamefont {A.~D.}\ \bibnamefont {Molodyk}}, \ and\ \bibinfo
  {author} {\bibfnamefont {S.~S.}\ \bibnamefont {Strelchenko}},\ }\enquote
  {\bibinfo {title} {Chemical bond in semiconductor and semimetal crystals},}\
  \ (\bibinfo  {publisher} {Nauka},\ \bibinfo {address} {Minsk},\ \bibinfo
  {year} {1973})\ p.\ \bibinfo {pages} {248}\BibitemShut {NoStop}%
\bibitem [{\citenamefont {Jain}\ \emph {et~al.}(2011)\citenamefont {Jain},
  \citenamefont {Hautier}, \citenamefont {Ong}, \citenamefont {Moore},
  \citenamefont {Fischer}, \citenamefont {Persson},\ and\ \citenamefont
  {Ceder}}]{PhysRevB.84.045115}%
  \BibitemOpen
  \bibfield  {author} {\bibinfo {author} {\bibfnamefont {A.}~\bibnamefont
  {Jain}}, \bibinfo {author} {\bibfnamefont {G.}~\bibnamefont {Hautier}},
  \bibinfo {author} {\bibfnamefont {S.~P.}\ \bibnamefont {Ong}}, \bibinfo
  {author} {\bibfnamefont {C.~J.}\ \bibnamefont {Moore}}, \bibinfo {author}
  {\bibfnamefont {C.~C.}\ \bibnamefont {Fischer}}, \bibinfo {author}
  {\bibfnamefont {K.~A.}\ \bibnamefont {Persson}}, \ and\ \bibinfo {author}
  {\bibfnamefont {G.}~\bibnamefont {Ceder}},\ }\href {\doibase
  10.1103/PhysRevB.84.045115} {\bibfield  {journal} {\bibinfo  {journal} {Phys.
  Rev. B}\ }\textbf {\bibinfo {volume} {84}},\ \bibinfo {pages} {045115}
  (\bibinfo {year} {2011})}\BibitemShut {NoStop}%
\bibitem [{\citenamefont {Hautier}\ \emph {et~al.}(2012)\citenamefont
  {Hautier}, \citenamefont {Ong}, \citenamefont {Jain}, \citenamefont {Moore},\
  and\ \citenamefont {Ceder}}]{PhysRevB.85.155208}%
  \BibitemOpen
  \bibfield  {author} {\bibinfo {author} {\bibfnamefont {G.}~\bibnamefont
  {Hautier}}, \bibinfo {author} {\bibfnamefont {S.~P.}\ \bibnamefont {Ong}},
  \bibinfo {author} {\bibfnamefont {A.}~\bibnamefont {Jain}}, \bibinfo {author}
  {\bibfnamefont {C.~J.}\ \bibnamefont {Moore}}, \ and\ \bibinfo {author}
  {\bibfnamefont {G.}~\bibnamefont {Ceder}},\ }\href {\doibase
  10.1103/PhysRevB.85.155208} {\bibfield  {journal} {\bibinfo  {journal} {Phys.
  Rev. B}\ }\textbf {\bibinfo {volume} {85}},\ \bibinfo {pages} {155208}
  (\bibinfo {year} {2012})}\BibitemShut {NoStop}%
\bibitem [{\citenamefont {Anderson}\ \emph {et~al.}(2003)\citenamefont
  {Anderson}, \citenamefont {Crisalle}, \citenamefont {Li},\ and\ \citenamefont
  {Holloway}}]{NREL_report}%
  \BibitemOpen
  \bibfield  {author} {\bibinfo {author} {\bibfnamefont {T.~J.}\ \bibnamefont
  {Anderson}}, \bibinfo {author} {\bibfnamefont {O.~D.}\ \bibnamefont
  {Crisalle}}, \bibinfo {author} {\bibfnamefont {S.~S.}\ \bibnamefont {Li}}, \
  and\ \bibinfo {author} {\bibfnamefont {P.~H.}\ \bibnamefont {Holloway}},\
  }\href@noop {} {\emph {\bibinfo {title} {Future CIS Manufacturing Technology
  Development}}},\ \bibinfo {type} {Tech. Rep.}\ (\bibinfo  {institution}
  {National Renewable Energy Laboratory},\ \bibinfo {address} {Golden,
  Colorado},\ \bibinfo {year} {2003})\BibitemShut {NoStop}%
\bibitem [{\citenamefont {Wagman}\ \emph {et~al.}(1982)\citenamefont {Wagman},
  \citenamefont {Evans}, \citenamefont {Parker}, \citenamefont {Schumm},
  \citenamefont {Halow}, \citenamefont {Bailey}, \citenamefont {Churney},\ and\
  \citenamefont {Nuttall}}]{NBS_tables}%
  \BibitemOpen
  \bibfield  {author} {\bibinfo {author} {\bibfnamefont {D.~D.}\ \bibnamefont
  {Wagman}}, \bibinfo {author} {\bibfnamefont {W.~H.}\ \bibnamefont {Evans}},
  \bibinfo {author} {\bibfnamefont {V.~B.}\ \bibnamefont {Parker}}, \bibinfo
  {author} {\bibfnamefont {R.~H.}\ \bibnamefont {Schumm}}, \bibinfo {author}
  {\bibfnamefont {I.}~\bibnamefont {Halow}}, \bibinfo {author} {\bibfnamefont
  {S.~M.}\ \bibnamefont {Bailey}}, \bibinfo {author} {\bibfnamefont {K.~L.}\
  \bibnamefont {Churney}}, \ and\ \bibinfo {author} {\bibfnamefont {R.~L.}\
  \bibnamefont {Nuttall}},\ }\href@noop {} {\emph {\bibinfo {title} {The NBS
  Tables of Chemical Thermodynamic Properties}}},\ \bibinfo {series} {Selected
  Values for Inorganic and C1 and C2 Organic Substances in SI Units},
  Vol.~\bibinfo {volume} {11}\ (\bibinfo  {publisher} {Amer Chemical Society
  and the American Institute of Physics for the National Bureau of Standards},\
  \bibinfo {address} {Washington, DC},\ \bibinfo {year} {1982})\BibitemShut
  {NoStop}%
\end{thebibliography}%

\end{document}